\newcommand{\Title}{Working Memory with Polychronous Chains}
\newcommand{\ShortTitle}{MNESIS: Memory with Polychronous Chains}
\newcommand{\Keywords}{spiking neural networks; polychronisation; heterogeneous delays; working memory.}
\newcommand{\FirstLP}{Laurent U}
\newcommand{\LastLP}{Perrinet}
\newcommand{\AuthorLP}{\FirstLP\ \LastLP}
\newcommand{\EmailLP}{laurent.perrinet@cnrs.fr} 

\newcommand{\Department}{Institut de Neurosciences de la Timone (UMR 7289)}%
\newcommand{\Affiliation}{Aix Marseille Univ, CNRS}%
\newcommand{\Street}{27 boulevard Jean Moulin}%
\newcommand{\PostCode}{13005}%
\newcommand{\City}{Marseille}%
\newcommand{\Country}{France}%
\newcommand{\WebsiteLP}{https://laurentperrinet.github.io}%
\newcommand{\Acknowledgements}{
This work was performed using HPC resources from GENCI-IDRIS (Grant 2025--AD010314955R2).
The authors declare no competing interests regarding this manuscript. The funders had no role in the design of the study; in the collection, analyses, or interpretation of data; in the writing of the manuscript; or in the decision to publish the results.
For the purpose of open access, the authors have applied a CC-BY public copyright licence to any Author-Accepted Manuscript version arising from this submission, and the different versions are available on arXiv at \url{https://arxiv.org/abs/2604.14096}.
}
\newcommand{\AIstatement}{
During the preparation of this paper, the Gemma4:31b generative AI tool (\url{https://ollama.com/library/gemma4:31b-mlx}) and Claude Opus 5.5 were used solely for language editing---improving the clarity, grammar, and phrasing of drafted text---and for limited technical assistance with code development. These tools were not used for generating research ideas, formulating research questions or hypotheses, designing the methodology, conducting or interpreting data analysis, or performing mathematical derivations or calculations. All intellectual contributions, analytical work, and mathematical reasoning in this paper are the authors' own. All AI-assisted content has been reviewed by the authors, who assume full responsibility for the accuracy and integrity of the entire work.
}
\newcommand{\Abstract}{%
Working memory --- the ability to store and recall precise temporal patterns of activity --- remains an open challenge for spiking neural networks (SNNs). A recurrent SNN of $N=1024$ neurons is proposed, in which each synapse is equipped with $D = 41$ heterogeneous delays, parameterised as a weight tensor $\mathbf{W} \in \mathbb{R}^{N \times N \times D}$ and trained end-to-end with surrogate-gradient backpropagation through time. Each stored motif is represented as a sequential chain of overlapping Spiking Motifs: contiguous context windows of length $D$ that uniquely predict the activity at the next time step. A closed-form Hebbian initialisation, derived by deconvolving the LIF membrane response and targeting a membrane potential value above the firing threshold $\vartheta$, already achieves high $F_1$ scores before any gradient step on a benchmark of $M=16$ motifs of duration $T=1000$ steps. With learning, the network reaches a mean $F_1$ score of $0.80$ over the motifs, and retrieval degrades gracefully when the trigger window is corrupted, with an $F_1$ score of $0.53$ at $25\%$ bit-flip noise. The network shows attractor-like retrieval dynamics consistent with hippocampal motif completion and with low-dimensional collective dynamics of population activity on manifolds. These results demonstrate that heterogeneous synaptic delays are an efficient and scalable substrate for working memory in SNNs, with direct implications for neuromorphic edge deployment. %
}

\documentclass{scrartcl}
\usepackage{fontspec}
\usepackage{geometry}
\usepackage{scrlayer-scrpage}
\clearpairofpagestyles
\usepackage{hyperref}
\hypersetup{colorlinks=true, linkcolor=blue, urlcolor=blue}

\usepackage{amsmath}
\usepackage{amsfonts}
\usepackage{amssymb}
\usepackage{graphicx}
\usepackage{booktabs}
\usepackage{nicefrac}
\usepackage{microtype}
\usepackage{xcolor}
\usepackage{adjustbox}
\usepackage{dsfont}
\usepackage{url}
\usepackage[style=numeric, sorting=none]{biblatex}
\newcommand{\citep}[1]{\parencite{#1}}
\newcommand{\citet}[1]{\textcite{#1}}
\graphicspath{{figures/}}
\title{\Title}
\author{
  \AuthorLP \\
  \texttt{\EmailLP} \\
  \Department \\
  \Affiliation \\
  \Street, \PostCode\ \City, \Country \\
  {\small\url{\WebsiteLP/publication/perrinet-26}}
}
\date{\today}

\begin{document}
\maketitle

\begin{abstract}
\Abstract
\end{abstract}
\section{Introduction}
\begin{figure}
  \includegraphics[width=\linewidth]{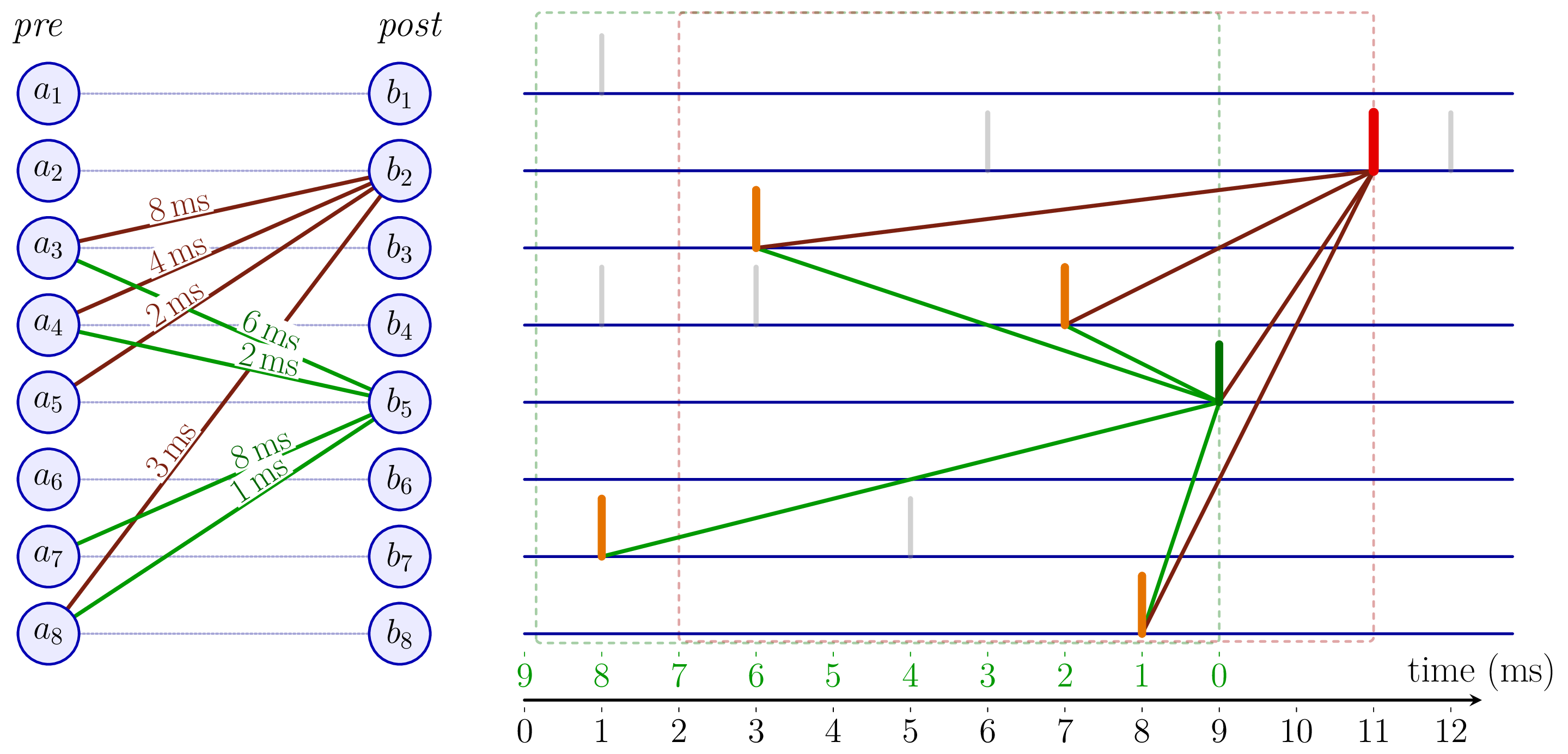}%
  \caption{%
    \textbf{Working memory as a chained sequential spike prediction via heterogeneous delays.}
    \emph{Left:} A recurrent HD-SNN of eight neurons numbered $1$--$8$ for which two subsets of recurrent connections are shown that define two Spiking Motifs from the pre-synaptic side (denoted $a$) to the post-synaptic side ($b$): dark-red connections from $a_2$, $a_3$, $a_4$, and $a_5$ target $b_1$ with delays $3$, $8$, $4$, $2$\,ms; green connections from $a_2$, $a_3$, $a_4$, and $a_5$ target $b_5$ with delays $1$, $6$, $2$, $8$\,ms. %
    \emph{Right:} Raster plot of the same two motifs: each trajectory joins an input spike to the spike it contributes to, and faint grey spikes are from spontaneous activity.
    \emph{Step~1, first prediction (green dashed box):} the orange input spikes emitted at staggered times by $a_2$--$a_5$ within the first context window (dashed green box) constitute a first Polychronous Group~\citep{Izhikevich2006}. As weights and delays can be tuned, this structure is called a Spiking Motif~\citep{Perrinet2023}. The green delays are tuned so that the four spikes, emitted at $t=8$, $3$, $7$ and $1$\,ms by $a_2$, $a_3$, $a_4$ and $a_5$, converge \emph{synchronously} at $b_5$ at $t=9$\,ms, crossing the firing threshold and producing the green output spike with a precise time.
    \emph{Step~2, second prediction (dark-red dashed box):} the green spike of $b_5$ re-enters the population as a new input of $a_5$. Together with the spikes still visible in the shifted context window (dark-red dashed box), spikes are routed through the dark-red delays in order to converge synchronously on $b_1$ at $t=11$\,ms, generating a new output spike.
    Because output spikes re-enter the same population, detecting one Spiking Motif provides the context for predicting the next spike in the sequence, implementing working memory as a chain of polychronous motifs in which each new temporal prediction is grounded in the preceding context window.
  }
  \label{fig:izhikevich}
\end{figure}

A century ago,~\citet{Adrian1926} established that neural communication relies on all-or-none action potentials, \emph{spikes}. Whether information is encoded in precise spike timing, for instance in coarsely synchronous activity with so-called \emph{synfire chains}, or in the coarser measure of instantaneous firing rate remains debated~\citep{Softky1993,Bohte2004,Grimaldi2023} --- a question that becomes especially sharp for working memory, where the brain must maintain and manipulate information across delays of seconds. In classical artificial neural networks, long-range temporal dependencies are addressed by gated recurrent unit (GRU) architectures or Transformers~\citep{Vaswani2017}, which decouple temporal binding range from the timescale of individual units. The analogous problem for spiking neural networks (SNNs) --- binding activity motifs separated by intervals far exceeding a single neuron's time constant --- lacks a general solution and remains a key challenge for biologically plausible sequence processing~\citep{Yin2026}.

\citet{Izhikevich2006} extended the synfire chains framework by introducing heterogeneous axonal delays in a recurrent population of spiking neurons trained with spike-timing-dependent plasticity (STDP). The resulting \emph{polychronous neuronal groups} (PNGs) --- time-locked, non-synchronous firing motifs at millisecond precision --- vastly outnumber the neurons in the network, yielding a large short-term memory capacity exploited for visual feature binding~\citep{Eguchi2018} and sequence classification~\citep{Paugam-Moisy2008}.~\citet{Szatmary2010} showed that short-term plasticity can trigger a PNG's reactivation and sustain a working memory. However, storing spike motifs of arbitrary length is inherently difficult: the sequential structure of the task means each context window is not an independent query but is itself the output of the network at the previous step, so small errors at early time steps propagate and amplify towards the end of the sequence --- a compounding difficulty absent from static associative memory like Hopfield networks.

The limitation of the original model from~\citet{Izhikevich2006} is that delays are fixed rather than learned. Recent work addresses this:~\citet{Goltz2021} demonstrated fast, energy-efficient inference on neuromorphic hardware using first-spike-time coding, while~\citet{Hammouamri2023} introduced gradient-based delay learning via dilated convolutions with learnable spacings (DCLS), achieving sparse solutions in which only $\sim\!20\%$ of delay channels remain active. Applied to event-based vision, heterogeneous learned delays in a single spiking layer suffice for fast motion detection~\citep{Grimaldi2023a}, scalable to always-on object recognition~\citep{Grimaldi2024}. Building on this, \emph{Spiking Motifs} (SMs) generalise PNGs to a gradient-trainable, hardware-compatible framework by replacing fixed STDP rules with end-to-end learned delays, here referred to as \emph{Heterogeneous-Delay SNNs} (HD-SNNs)~\citep{Perrinet2023}. However, these motifs have a short time span equal to the range of possible delays, whereas stimulation experiments in mice show that sequences lasting seconds influence neural activity~\citep{Deveau2026}.~\citet{Kronland-Martinet2026} recently extended this to motifs of arbitrary length by chaining sub-motif detectors with bounded synaptic delays, achieving robust encoding of complete Spiking Heidelberg Digits motifs.

The present work extends this chain-of-motifs framework~\citep{Kronland-Martinet2026} to a fully recurrent topology applied to modelling working memory. Indeed, such a memory has been shown to be essential in recent models including a predictive architecture~\citep{Dawid2023}, yet a spiking version of the predictive module is still missing. Rather than relying on fixed delays and unsupervised STDP as in the original polychronisation model, the network is trained end-to-end with surrogate gradients~\citep{Neftci2019a}: once primed by a short trigger window of $D$ steps (the maximum synaptic delay, Section~\ref{sec:methods}), it autonomously recalls arbitrary-length spike sequences, bridging polychronisation theory with gradient-based machine learning.

\section{Methods}
\label{sec:methods}
\subsection{MNESIS, a Recurrent HD-SNN architecture} 

In a previous work, a feedforward HD-SNN was shown to learn to detect Spiking Motifs~\citep{Perrinet2023}. However, this architecture is limited to motifs of length $D$, the maximum synaptic delay. To store and recall longer sequences, it is possible to extend this architecture to a fully recurrent topology, where each neuron receives inputs from all other neurons across all delays. The core concept is to retrieve a memory progressively: each motif of length $D$ predicts the next spikes in the memory to retrieve. This allows the network to generate a chain of overlapping Spiking Motifs, where each motif provides the context for predicting the next spikes in the sequence (see Figure~\ref{fig:izhikevich}).

In practice, the study focuses on MNESIS, a recurrent HD-SNN which consists of a population $\mathcal{P}$ of $N = 1024$ leaky integrate-and-fire (LIF) neurons implemented in \texttt{snnTorch}~\citep{Eshraghian2023} and simulated over $T = 1000$ steps of $1\,\mathrm{ms}$ each (total duration: $1\,\mathrm{s}$). Extending the feedforward Spiking Motif Detector~\citep{Perrinet2023}, the architecture includes full recurrence: each neuron $j \in \mathcal{P}$  integrates inputs from \emph{all} other neurons across \emph{all} preceding $D = 41$ time steps. Such a network is the most general for learning purposes, and at inference, pruning of the synapses may be added to gain from the sparseness of synapses on the energy budget. The maximum delay $D$ is justified by biological observations and is of the same order of magnitude as in the original polychronisation study~\citep{Izhikevich2006}. For each neuron $j \in \mathcal{P}$, the membrane potential $u_j(t)$ evolves as
\begin{equation}
  u_j(t)
  = \beta \cdot u_j(t-1) \cdot (1 - s_j(t-1))
    + \sum_{i \in \mathcal{P}}
      \sum_{d=1}^{D} \mathbf{W}_{j, i, d} \cdot s_i(t-d) ,
  \label{eq:lif}
\end{equation}
where $\beta \in (0,1)$ is the membrane decay constant ($\beta = 0.7$ in the simulations, that is, a time constant of $\tau= 1 / \ln(\frac 1 \beta) \approx 2.8\,\mathrm{ms}$), $\mathbf{W}_{j, i, d}$ is the synaptic weight to neuron $j \in \mathcal{P}$ from neuron $i \in \mathcal{P}$ at delay $d$, and $s_j(t) \in \{0,1\}$  is the spike indicator $s_j(t) = \mathds{1}[u_j(t) \geq \vartheta]$, that is, $s_j(t)=0$ if $u_j(t) < \vartheta$ and a spike is emitted $s_j(t)=1$ when $u_j(t)$ exceeds the threshold $\vartheta$. After spiking, the membrane potential that is accumulated in Equation~\ref{eq:lif} is reset to zero.

\begin{figure}
  \centering
  \adjustbox{valign=c,width=0.49\textwidth}{%
  \includegraphics[width=\linewidth]{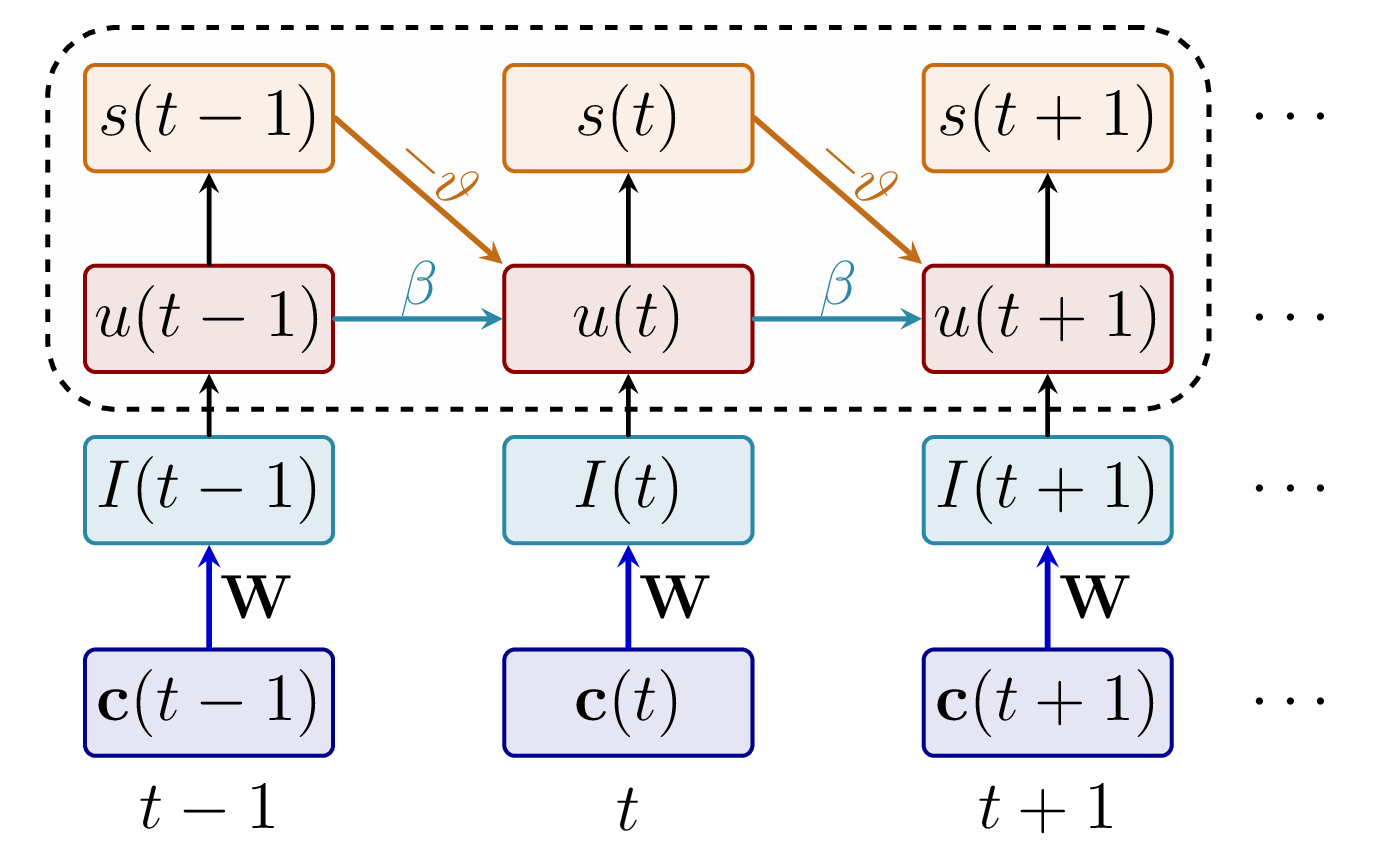}%
  }
    \hfill
    \adjustbox{valign=c}{\includegraphics[width=0.49\textwidth]{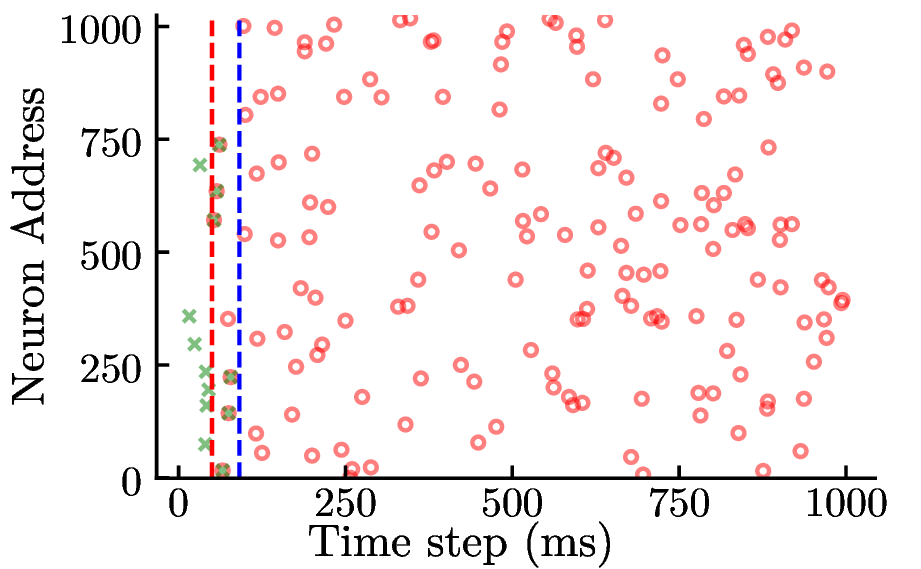}}
  \caption{%
    Network architecture and a sample target motif. \emph{Left:}
    Unrolled view of the recurrent HD-SNN over time (one column representing one time step), illustrating delayed connections between neurons: Blue arrows indicate synaptic connections that integrate past activity across heterogeneous delays: the arrow labelled $\mathbf{W}$ stands for the weights over all $D$ delays. The context $\mathbf{c}(t)$ is the input received by the network at time $t$: the spike history over the $D$ previous steps, that is, the activity from $t-D$ to $t-1$ included ($t$ excluded), made of the target motif clamped during the trigger window, then the network's own recurrent spikes. Clamping input spikes is what excites the network: the trigger window is one such case, and other input spikes can be clamped in the same way. Once the trigger window is over, $\mathbf{c}(t)$ contains the spikes $s(t-1)$ emitted by the network itself, so that its output feeds back onto its input: this closes a recurrence loop, $s(t) \to \mathbf{c}(t+1) \to I(t+1) \to u(t+1) \to s(t+1)$, which the figure does not draw. The arrow labelled $-\vartheta$ depicts the subtractive reset of snnTorch, which subtracts the threshold from the membrane after each spike; the simulations reset the membrane to zero instead (Equation~\ref{eq:lif}), and the subtractive reset was one of the options scanned in Section~\ref{sec:hyperparameters}.
    \emph{Right:}
    One representative target spike motif, the second of the $M=16$ motifs, also shown in Figure~\ref{fig:target} ($N=1024$ neurons, $T=1000$ time steps); each red circle marks one target spike. The goal of the network is to memorise $M=16$ such motifs. Each trial is preceded by $50$ time steps of spontaneous background activity (random spikes at rate $p_A$) before the stimulus. The trigger window (green crosses between the dashed lines) consists of the first $D=41$ time steps of the target motif clamped to the network input; the recurrent network then autonomously predicts the remaining spikes.
  }\label{fig:model}
\end{figure}

In the implementation, the spike history of all $N$ neurons over the past $D$ steps is concatenated into a flattened context vector of dimension $N \cdot D$ at each time step, so that $\mathbf{W}$ is reshaped as $\mathbf{\bar W} \in \mathbb{R}^{N \times (N \cdot D)}$. This is equivalent to the temporal convolution used previously in a feedforward setting~\citep{Perrinet2023}, here extended to a fully recurrent topology. The HD-SNN thus maps onto a standard static computational graph that can be trained with surrogate gradient descent as classical SNNs (see Figure~\ref{fig:model}, left panel).

\subsection{Synthetic dataset and definition of noise}
To assess working-memory capacity in all generality, a dataset of $M = 16$ distinct target motifs $S^{(\mu)} \in \{0,1\}^{N \times T}$ was generated. Following previous work~\citep{Perrinet2023}, each motif is generated using a stochastic spiking pattern generator with both a structured and background component. Specifically, for each motif, a \emph{frozen} motif $P^{(\mu)} \in \{0,1\}^{N \times T}$ is drawn with $P^{(\mu)} \sim \mathrm{Bernoulli}(p_A)$, where $p_A = 1.6 \times 10^{-4}$ is the average spike probability per neuron and per time step ($0.16\,\mathrm{Hz}$ for $1\,\mathrm{ms}$ time bins). This rate follows the estimate of the average spiking frequency in the human brain~\citep{Lennie2003}.

To account for the stochasticity inherent to neural activity, a balanced bit flipping mechanism is further included with probability $p_\mathrm{flip} = 0.01$. This is defined by the binary mask  $B^{(\mu)} \in \{0,1\}^{N \times T}$ of bit flips locations:
$
  B^{(\mu)} \sim \mathrm{Bernoulli}(p_\mathrm{flip})
  $.
Values are left untouched outside the mask and will be drawn with probability $p_A$ on the mask. In other words, the spike motifs $S^{(\mu)} \in \{0,1\}^{N \times T}$ are those of the frozen motif $P^{(\mu)}$ except where $B^{(\mu)} = 1$ for which they are obtained by drawing with probability $p_A$, that is $R^{(\mu)} \sim \mathrm{Bernoulli}(p_A)$. Formally, this can be expressed as:
\begin{equation}
  S^{(\mu)} = P^{(\mu)} \cdot (1 - B^{(\mu)}) + R^{(\mu)} \cdot B^{(\mu)}
  \label{eq:pattern_generation}
\end{equation}
Note that the bit flipping mechanism introduces a stochasticity into the structure of the spike motifs, however, it does not change the average firing rate. Indeed, since the generation of $P^{(\mu)}$ and $B^{(\mu)}$ are independent, the expected firing rate $E(S^{(\mu)})$ is equal to $p_A \cdot (1 - p_\mathrm{flip}) + p_A \cdot p_\mathrm{flip} = p_A$.
%
\subsection{Working memory task: memory retrieval}
\label{sec:task}
%
Given a set of target motifs as defined above, a generic working memory task consists of training the network to reproduce the target motif that follows a short \emph{trigger} window. Formally, given $S^{(\mu)}$, each contiguous window of length $D$, i.e.\ the slice $S^{(\mu)}[\cdot,\, t{-}D:t{-}1]$, constitutes a unique context that may be used to predict the spiking activity $s_j(t)$ of all neurons $j \in \mathcal{P}$ at time $t$ (see Figure~\ref{fig:izhikevich}). The network is therefore trained to generate a sequential chain of overlapping Spiking Motifs~\citep{Perrinet2023,Kronland-Martinet2026} that faithfully reproduces the target. Memory retrieval is triggered by clamping all neuron activities to the target motif during the trigger window which corresponds to the initial $D$ time steps ($0 \le t < D$), that is, corresponding to one elementary Spiking Motif (see Figure~\ref{fig:model}, right panel); the goal of the network is then to generate the remainder of the sequence autonomously.

This chain of contexts has a structural weak point. A motif may contain a \emph{gap}, that is, a stretch of $D$ consecutive time steps in which none of the $N$ neurons fires. The context window that covers the gap is then empty, and since the synaptic current $I_j(t) = \sum_{i,d} \mathbf{W}_{j,i,d} \cdot s_i(t-d)$ has no bias term, it vanishes whatever the weights: no readout of the past $D$ steps can predict the spikes that follow the gap, and an autonomous replay necessarily stops at its onset. A given window is empty with probability $(1-p_A)^{N D} \approx e^{-p_A N D}$, that is $\approx 1.2 \times 10^{-3}$ for the default parameters ($p_A N D \approx 6.7$). Counting the gaps that open just after a time step containing at least one spike gives an expected number of $(T-D) \cdot (1 - e^{-p_A N}) \cdot e^{-p_A N D} \approx 0.18$ gaps per motif, so that about one motif in six is expected to contain at least one. The gap probability is kept as it is, to avoid introducing heuristics; the ability of the network to retrieve motifs in the presence of gaps is shown in Section~\ref{sec:results}.

As an accuracy metric, the $F_1$ score is used between the predicted spike trains and the frozen motif $P^{(\mu)}$, whatever the realisation used to build the trigger, computed on the true and false positives and negatives pooled over time (in the prediction window), neurons and motifs. The $F_1$ score is the harmonic mean of precision (proportion of predicted activity that is correct) and recall (proportion of the target activity that is successfully reproduced). Knowing a priori that average neural activity is sparse, maximising $F_1$ simultaneously penalises over-active networks (low precision) and silent networks (low recall).

Applied to the synthetic motifs, knowing $p_\mathrm{flip}$ then one may first compute the precision as the ratio of expected true positives to the sum of true positives and false positives, that is, $\frac{p_A \cdot (1 - p_\mathrm{flip} + p_A \cdot p_\mathrm{flip})}{p_A \cdot (1 - p_\mathrm{flip} + p_A \cdot p_\mathrm{flip}) + p_\mathrm{flip} \cdot (1 - p_A) \cdot p_A} = 1 - p_\mathrm{flip} \cdot (1 - p_A)$. Then, recall is equal to precision as the average ratio of false negatives $p_A \cdot p_\mathrm{flip} \cdot (1 - p_A)$ is equal to that of false positives (remember that the average frequency is not changed by the balanced bit flip operation). As a result, the average $F_1$ score between the frozen and perturbed motifs equals the precision and recall and is given by:
\begin{equation}
  F_1 = 1 - p_\mathrm{flip} \cdot (1 - p_A)
  \label{eq:f1}
\end{equation}
This value is the score of a noisy copy. It bounds the $F_1$ score that can be reached when the output is scored against a noisy realisation, as during training, since the flipped bits of a realisation cannot be predicted. It is not, however, a bound for retrieval, which is scored against the frozen motif $P^{(\mu)}$: a network that has stored the frozen motif itself, rather than one of its realisations, can replay it exactly from a noisy trigger window, and thus exceed this value (see Section~\ref{sec:results}).

\subsection{Analytical weight initialisation}
\label{sec:init}
From Equation~\ref{eq:lif}, by normalising the potential to arbitrary units, the membrane potential is required to reach a sub-threshold value close to $1$ whenever a spiking motif from a target motif occurs. Before training, the weight tensor $\mathbf{W}$ can be initialised analytically by treating this task as a linear regression problem. For each target motif $S^{(\mu)} \in \{0,1\}^{N \times T}$, taken here as the frozen motif $P^{(\mu)}$ (that is, with $p_\mathrm{flip} = 0$), the context window at each time point $t$ is the flattened vector $\mathbf{c}^{(\mu)}(t) \in \{0,1\}^{N \cdot D}$ containing the spike history of all neurons over steps $t-D$ to $t-1$ and the corresponding desired output $\mathbf{Y}(t) \in \{0,1\}^N$. Stacking all $Q = M \cdot (T-D)$ context--output pairs across motifs and time steps yields
\[
  \mathbf{C} \in \{0,1\}^{(N \cdot D) \times Q},
  \qquad
  \mathbf{Y}^* \in \mathbb{R}^{N \times Q},
\]
and the initialisation problem becomes $\mathbf{W}\,\mathbf{C} \approx \mathbf{Y}$.

The choice of regression output $\mathbf{Y}$ requires care. A naive choice is the binary spike matrix $S$ itself, which would amount to ask the network to produce a current equal to the target spike at each step. However, the synaptic current $I_j(t) = \sum_{i,d} \mathbf{W}_{j,i,d} \cdot s_i(t-d)$ is not compared directly to $s(t)$: it is first \emph{filtered} by the membrane, which acts as a causal first-order IIR lowpass with transfer function $H(z) = 1/(1 - \beta \cdot z^{-1})$ and impulse response $h(t) = \beta^t$. The correct output is therefore obtained by deconvolution.  The inverse filter of $H(z)$ is $H^{-1}(z) = 1 - \beta \cdot z^{-1}$. Applied to a target trajectory that reaches $1$ at $t$ and resets immediately, the optimal input current is therefore for each neuron $j \in \mathcal{P}$ equal to:
\begin{equation}
  I_j(t) \;=\; s_j(t) - \beta \cdot s_j(t-1).
  \label{eq:deconv_target}
\end{equation}
This is a \emph{biphasic} waveform: $1$ at each target spike time $t$ and $-\beta$ exactly one step later, cancelling the residual membrane charge left by the leak. For $\beta \to 0$ the biphasic correction collapses to a scaled impulse; for $\beta = 0.7$ the negative lobe is substantial and significantly reduces spurious spikes.

Since the system is underdetermined ($Q < N \cdot D$), the minimum-$\ell_2$-norm solution is the Moore--Penrose pseudo-inverse $\mathbf{W}^* = \mathbf{Y}^* \cdot \mathbf{C}^+$. With sparse neural activity at rate $p_A = 1.6\cdot 10^{-4}$, consecutive context windows are nearly orthogonal, so the Gram matrix of the contexts satisfies $\mathbf{C}^\top\mathbf{C} \approx N \cdot D \cdot p_A \cdot \mathrm{Id}$, giving the closed-form initialisation for each neuron pair $i, j \in \mathcal{P}$ and delay $d \in \{1, \ldots, D\}$:
\begin{equation}
  w_{i, j, d}
  = \frac{1}{N \cdot D \cdot p_A}
    \sum_{\mu=1}^{M}\sum_{t=D+1}^{T}
    s_i^{(\mu)}(t-d)
    \cdot
    \bigl(s_j^{(\mu)}(t) - \beta \cdot s_j^{(\mu)}(t-1)\bigr).
  \label{eq:init}
\end{equation}
Each weight $w_{i, j, d}$ is proportional to the cross-correlation between the deconvolved spike train of neuron $j$ and the raw spike train of neuron $i$ at delay $d$, summed over all $M$ motifs and normalised by the mean number $N \cdot D \cdot p_A$ of spikes in a context.

The threshold is set at $\vartheta = 0.75$ at each target spike time $t$, reserving a safety margin $\delta = 1 - \vartheta = 0.25$. This margin prevents spurious firings from partial contexts and places the membrane in the regime where the surrogate gradient is near its maximum, so that the first gradient step is maximally informative. A perfect context thus drives the membrane to $u_j(t^*) = 1$, a margin $\delta = 0.25$ above the threshold, so that a context missing up to a quarter of its supporting spikes still evokes the target spike. Note that at inference time, the network receives its own generated context rather than the ground truth, so errors can accumulate over time --- the primary source of difficulty analysed in Section~\ref{sec:results}.
\subsection{Training}
The loss is defined as $\mathcal{L} = 1 - F_1$, such that minimising this loss (with a maximum of $1$ and minimum of $0$) maximises the mean $F_1$ score across all $M$ target motifs. Learning is carried out with surrogate-gradient backpropagation through time (BPTT)~\citep{Neftci2019a}, using a fast-sigmoid surrogate with sharpness parameter $\alpha = 5$. Weights are optimised with Adadelta, using a cosine learning-rate schedule with warmup, as used for instance in~\citep{Hammouamri2023}, from a base learning rate $\eta_\mathrm{base} = 40.0 \times 10^{-3}$ to a final learning rate $\eta_\mathrm{final} = 400 \times 10^{-6}$, with $16$ warmup epochs over $256$ total epochs. At each epoch, a new stochastic realisation of the motifs (Equation~\ref{eq:pattern_generation}) is generated, acting as data augmentation. Dropout regularisation with probability $0.60$ is applied to the post-synaptic current. The validation loss, computed for another realisation used as the trigger and scored against the frozen motifs $P^{(\mu)}$, is monitored for detecting overfitting; all realisations are drawn from explicit seeds, so that a run is reproducible.

The gradient-based learning in the recurrent HD-SNN shares fundamental principles with biological spike-timing-dependent plasticity (STDP)~\citep{Izhikevich2006} and classical spike-based Hebbian learning~\citep{Perrinet2002}. Both mechanisms exploit temporal correlations in spike timing to adjust synaptic strengths. However, while STDP implements local pairwise updates based on individual spike timings, the present training performs global optimisation that effectively learns the collective interaction of multiple delayed spike trains across the entire network. The analytical initialisation (see Equation~\ref{eq:init}) reveals this connection most clearly: it implements a Hebbian-like rule that resembles an average of STDP updates over all stored motifs, suggesting that gradient descent in this context can be viewed as an implementation of temporally structured synaptic plasticity.

The implementation of the MNESIS recurrent HD-SNN is available at \url{https://github.com/SpikeAI/MNESIS}. It was implemented using the snnTorch~\citep{Eshraghian2023} library with custom extensions for heterogeneous delay handling. The implementation leverages PyTorch with the Metal Performance Shader (MPS) backend for accelerated computation on Apple Silicon (M3 Ultra chip featuring 32 CPU cores and 80 GPU cores, with 512\,GB of unified memory architecture) and further tested with the CUDA backend on the Jean Zay supercluster at GENCI\@. The MPS and CUDA backends enabled efficient training and simulation of the large recurrent networks required for working memory tasks, leveraging the parallel processing capabilities of the GPU cores while maintaining low-latency access to the substantial memory resources necessary for storing all tensors' parameters. Training typically converged within $256$ epochs, requiring approximately $25$ minutes of computation time per training.
%
\section{Results}
\label{sec:results}
\subsection{Training dynamics and results}
\begin{figure}[!ht]
  \begin{center}
    \includegraphics[width=\textwidth]{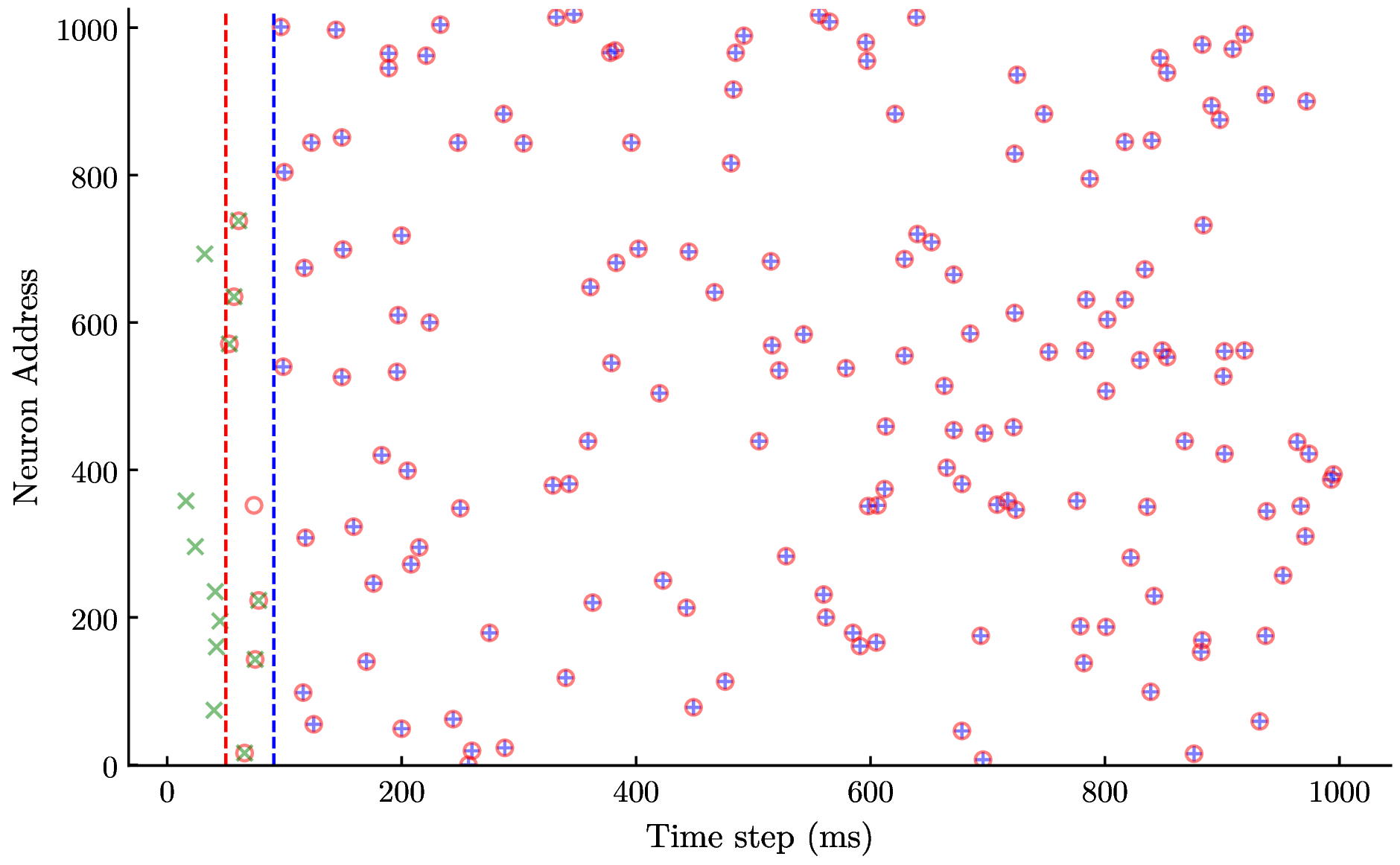}
  \end{center}
   \caption{Pattern retrieval by the recurrent HD-SNN after training. The network is clamped with input spikes (green crosses) comprising $50$ steps of spontaneous activity (left of the red dashed line) followed by the trigger window consisting of the initial sequence of one of the $M=16$ target motifs ($0 \leq t < D = 41$, delimited by the red and blue dashed lines). The motif shown, the second one, was chosen as characteristic of the average behaviour: with an intact trigger window, its $F_1$ score ($\approx 0.96$) is close to the median over the $M=16$ motifs ($\approx 0.95$). The autonomous replay is then simulated over the full sequence duration ($T = 1000$ steps). The raster plot shows network activity (blue plus signs, $N=1024$ neurons) which accurately matches the hidden target motif which was triggered during the trigger window. Note that the target spikes after the trigger window (red circles right of the blue dashed line) are hidden from the network.}\label{fig:target}
\end{figure}

%
The MNESIS recurrent HD-SNN was trained with surrogate-gradient BPTT~\citep{Neftci2019a}, starting from the analytical initialisation, on $M=16$ target motifs with $N=1024$ neurons, $D=41$ delays, $T=1000$ time steps, $p_A=1.6\times10^{-4}$, and $\beta=0.7$ (time constant $\tau \approx 2.8\,\mathrm{ms}$). Without the analytical initialisation, two characteristic failure modes arise during early training: episodes of complete silence, in which the network emits no spikes, and over-activation, in which nearly all neurons fire at every time step. Both are well-known pathological attractors of recurrent SNNs~\citep{Izhikevich2006,Voges10jpp}. The $F_1$ score is the harmonic mean of precision and recall, and maximising it therefore penalises both failure modes symmetrically: low precision corresponds to an over-active network and low recall to a silent one. The cosine learning-rate schedule~\citep{Hammouamri2023} helped guide the network away from these extremes towards a balanced, sparse-firing regime corresponding to that of the desired motifs. Training refines the analytical initialisation only slightly: the trained readout differs from it by $0.17\%$ in relative norm.

The analytical weight initialisation (Equation~\ref{eq:init}) provides the starting point of training. The Gram-matrix approximation $\mathbf{C}^\top\mathbf{C} \approx N \cdot D \cdot p_A \cdot \mathrm{Id}$ reduces the pseudo-inverse to a simple cross-correlation (Equation~\ref{eq:init}), but it does not preserve the accuracy. It assumes that every context contains the same number of spikes, whereas this number fluctuates around $N D p_A \approx 6.7$: with a normalisation by $N D p_A$, the sparsest contexts fall below threshold and the replay dies out ($F_1 \approx 0.12$ for the approximate initialisation), whereas a smaller normalisation makes it explode. The exact pseudo-inverse, whose weights adapt to the number of spikes of each context, is therefore used throughout, and training starts from it ($F_1 \approx 0.80$ over the $M=16$ motifs, Section~\ref{sec:results}). 

Computing the mean $F_1$ score, however, hides a strong heterogeneity between motifs. Scored motif by motif, for the predicted spike trains produced by the network triggered by a noisy realisation of the trigger window ($p_\mathrm{flip} = 0.01$) and preceded by $50$ steps of spontaneous activity, $9$ of the $M = 16$ motifs were replayed exactly, without a single false positive or false negative ($F_1 = 1$). This exceeds the value $F_1 \approx 0.990$ of Equation~\ref{eq:f1} for a noisy copy: the replay restores the frozen motif, not the noisy realisation that provided the trigger window. The other seven motifs failed in one of two ways, which correspond to the two ends of the distribution of the number $K$ of spikes in a context window.

Four motifs (the 4th, 10th, 12th and 13th) stopped after a given time step, their replay ending at a gap of $D$ silent steps (recall from $0.18$ to $0.56$). These are exactly the four motifs that contain such a gap ($K = 0$, Section~\ref{sec:task}), and the last correctly evoked spike immediately precedes the onset of the gap. This failure is structural rather than a defect of the weights: when the readout is fed with the exact history of the motif instead of the network's own spikes, the $F_1$ score of every motif, these four included, lies between $0.99$ and $1$, and the same four motifs failed in every realisation tested.

Three motifs (the 1st, 5th and 11th) were over-active instead, with a recall of $1$ and a precision from $0.55$ to $0.65$. The 13th motif, which also stops at its gap, is over-active too (precision $0.46$). Their false positives were not diffuse: all but $8$ of the $425$ false positives belonged to another stored motif, and the remaining ones to the same motif shifted in time, so that a second memory was replayed with a time shift alongside the correct one. As the readout is linear and the contexts nearly orthogonal, both chains should coexist without disrupting each other. With the minimum-norm weights, each of the $K$ spikes that support a target spike carries a weight of about $1/K$, so that a context with few spikes ($K \le 3$ for $11\%$ of the contexts) can be completed by a handful of spurious spikes and ignite the corresponding motif. Such intrusions are seeded before the replay starts, while the network already runs on the spontaneous activity that precedes the trigger window: without this activity, only one intrusion remained. Across realisations, the motif whose intrusion is driven by its trigger window alone was affected each time, whereas the other motifs without a gap were degraded only occasionally (and, when they were, mostly over-active), and the mean $F_1$ score ranged from $0.79$ to $0.84$ ($0.96$ when the four motifs with a gap are excluded). The gaps thus set a deterministic ceiling on the mean $F_1$ score, whereas intrusions make it vary from one realisation to the next.

\subsection{Retrieval of the different spike motifs}
\begin{figure}[!ht]
\begin{center}
\includegraphics[width=.92\textwidth]{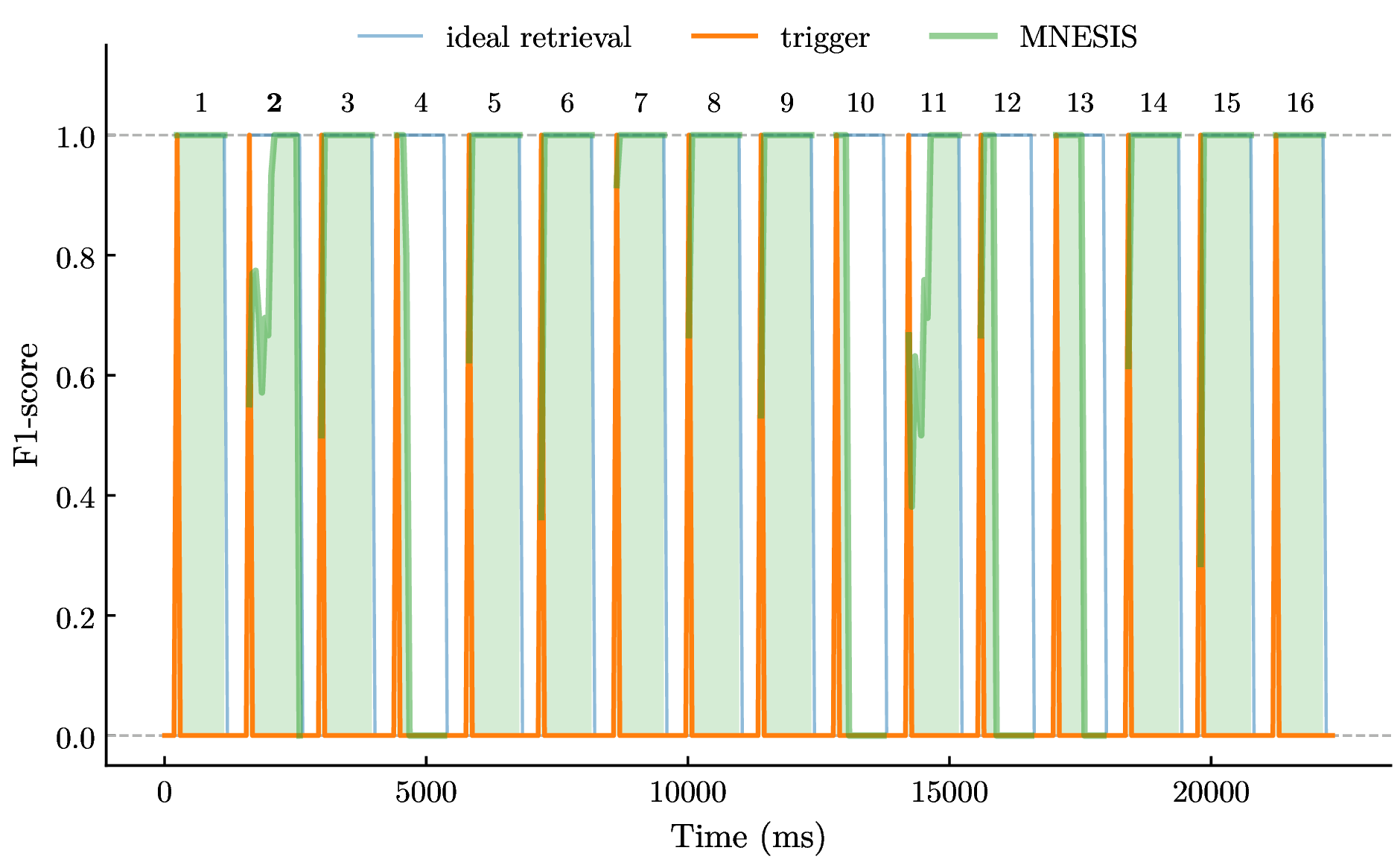}
\end{center}
\caption{Sequential retrieval of the $M=16$ stored motifs. The motifs are concatenated into a single stream, each flanked on both sides by $200$ silent steps, and only their first $D=41$ steps are given to the network as a trigger. The $F_1$ score (green), computed in consecutive $60$-step windows against the frozen motifs, is shown over the duration of each motif (it is not shown in the silent intervals, where the target is empty), together with the trigger mask, equal to $1$ on each trigger window (orange), and the ideal retrieval envelope, equal to $1$ over the full $T=1000$-step duration of each motif (blue). The motifs are numbered above their envelope, and the number of the second one, shown in Figure~\ref{fig:target}, is in bold. Retrieval succeeds for $12$ of the $16$ motifs: the $F_1$ score rises at the trigger onset and stays close to $1$ until the end of the motif, in two cases (2nd and 11th motifs) after a partial dip that follows the trigger. For the 4th, 10th, 12th and 13th motifs, the replay starts correctly but breaks down after about $200$ to $600$ steps: the $F_1$ score then collapses to $0$ and does not recover before the end of the motif. Grey dashed lines indicate extremal $F_1$ values $0$ or $1$.}\label{fig:retrieval}
\end{figure}


Figure~\ref{fig:target} shows the replay of the second motif, which matches its hidden target over the whole sequence. When the replay errs, the error follows the sequential structure of the task: because each context window $[t-D,\, t-1]$ uniquely predicts the activity at time $t$, errors at early time steps propagate into all subsequent predictions.

To evaluate the network's ability to retrieve and distinguish multiple stored motifs in sequence, all $M=16$ motifs were concatenated into a single long raster, each trial flanked on both sides by $200$ silent steps, without spontaneous activity, so as to isolate trigger-driven retrieval (Figure~\ref{fig:retrieval}). Each motif was triggered by its individual trigger window, and retrieval was scored by the $F_1$ score in consecutive $60$-step windows, against the frozen motifs. This is one realisation (seed $2026$), so no confidence interval is given. For $12$ of the $16$ motifs, the $F_1$ score rose at the onset of the trigger window and stayed high until the end of the motif (mean over its windows $0.96$, between $0.83$ and $1.00$): a single trigger window then suffices to select and replay the corresponding stored motif. For the four remaining motifs (the 4th, 10th, 12th and 13th), the replay started correctly but broke down after about $200$ to $600$ steps, and the $F_1$ score then stayed at $0$ until the end of the motif. Over the $M=16$ motifs, the mean $F_1$ score is $0.80$. This abrupt and irreversible failure is the error propagation of the training: once the activity deviates from the stored motif, the following context windows no longer correspond to it and the replay cannot recover. The three motifs that are over-active under the protocol of the training analysis (1st, 5th and 11th) are not over-active here, since the stream contains neither spontaneous activity nor bit flips: their $F_1$ scores are $1.00$, $0.98$ and $0.83$.

These failures are predicted, with no fitted parameter, by the cascade model of Section~\ref{sec:robustness}, which treats the replay as a positive feedback loop decided at its first evoked spike: a motif either ignites and is replayed in full, or it stops at its first gap. A spike of a motif is evoked only if the $D$ preceding steps of that motif hold at least one of its spikes ($K \geq 1$), so that a gap of $D$ silent steps stops the replay, whatever the trigger window. For random motifs of the same statistics, such a gap occurs with probability $\approx 0.17$ (estimated on $1000$ motifs), that is, in $2.8$ of $16$ motifs on average. Simulated step by step on the $16$ stored motifs, the cascade model predicts that exactly the 4th, 10th, 12th and 13th motifs stop, at the onset of their first gap, $233$, $258$, $268$ and $568$ steps after the onset of the motif, consistent within the $60$-step resolution of the windows with the breakdowns of Figure~\ref{fig:retrieval}, whose last window with $F_1 \geq 0.5$ ends at $280$, $280$, $300$ and $580$ steps. Under the protocol of the training analysis (noisy trigger, $50$ steps of spontaneous activity), it also predicts the recall of each motif, $0.175$, $0.221$, $0.294$ and $0.564$ for these four motifs and $1$ for the twelve others, equal to the measured values to three decimals, hence the pooled recall $r_0 = 0.825$ of the network. The model does not predict the precision ($0.835$), which is lowered by the over-active replays of the 1st, 5th, 11th and 13th motifs described above: they arise from the crosstalk between stored motifs, which the mean-field model leaves out.

\subsection{Effect of hyperparameters}
\label{sec:hyperparameters}
The learning hyperparameters were explored jointly with a Bayesian optimisation (Optuna~\citep{Akiba2019}, multivariate tree-structured Parzen estimator) over $200$ trials of $128$ epochs each, that is, half of the production budget. Of these, $197$ trials were completed; three were still marked as running in the study database. Each trial sampled the optimiser, the LIF parameters ($\beta$, $\vartheta$, reset mechanism), the surrogate sharpness $\alpha$, the learning-rate schedule and the dropout rate, and was scored by the loss $\mathcal{L} = 1 - F_1$ on fresh realisations of the motifs. Parameter importances were estimated with functional ANOVA (fANOVA). Three parameters dominated, together accounting for $0.88$ of the total importance: the firing threshold $\vartheta$ ($0.52$), the optimiser ($0.20$) and the pseudo-inverse initialisation ($0.15$). Their effect on the loss is reported first, then that of the remaining parameters in decreasing order of importance.

\paragraph{Firing threshold $\vartheta$.}
The threshold was sampled log-uniformly over a factor $3.618$ around its default value, that is, in $[0.21, 2.7]$. Among the Adadelta trials, the median loss was $\mathcal{L} \approx 0.82$ for $\vartheta \le 0.4$, fell to between $0.14$ and $0.17$ for $0.5 < \vartheta \le 0.8$, rose slightly to between $0.19$ and $0.23$ for $0.8 < \vartheta \le 1$, and every one of the $24$ trials with $\vartheta > 1$ failed (median loss $1$, best loss $0.70$), whether or not $\vartheta$ was learnt. The best $10\%$ of trials all had $\vartheta \in [0.59, 0.78]$ (median $0.68$), a range that includes the default value $\vartheta = 0.75$. Both failure modes follow from the analytical initialisation (Section~\ref{sec:init}), which drives the membrane to $1$ for a perfect context: above $\vartheta = 1$ the initialised network evokes almost no spike, and training did not recover from this silent state; well below it, the margin $\delta = 1 - \vartheta$ becomes so large that partial contexts and residual charges also cross the threshold, producing spurious spikes.

\paragraph{Optimiser.}
The importance of the optimiser reflects the failure of three of the five candidates rather than a fine ranking: Adam, AdamW and RMSprop reached a median loss $\mathcal{L} \approx 1$ (best losses $0.999$, $0.30$ and $0.999$ respectively), whereas Adadelta (median $0.26$, best $0.087$) and SGD with momentum (median $0.24$, best $0.16$) both learnt. Adadelta reached the lowest losses, and all of the best $10\%$ of trials used it. The momentum coefficients of the failing optimisers were however derived from the same search ranges as the decay rate of Adadelta, giving values above $0.989$ instead of the usual $0.9$, so that their failure may reflect this parameterisation rather than the optimisers themselves. Within Adadelta, the decay rate of the running averages (importance $0.04$) had no monotonic effect.

\paragraph{Pseudo-inverse initialisation.}
Replacing the pseudo-inverse $\mathbf{W}^* = \mathbf{Y}^* \cdot \mathbf{C}^+$ by its Hebbian approximation (Equation~\ref{eq:init}) prevented learning: among the Adadelta trials, the median loss was $\approx 1$ with the Hebbian rule (best loss $0.62$) against $0.22$ with the pseudo-inverse (best loss $0.087$), and the best $10\%$ of trials all used the pseudo-inverse. With $N \cdot D \cdot p_A \approx 6.7$ active entries per context, the contexts are only approximately orthogonal, and the exact pseudo-inverse plausibly removes the crosstalk between overlapping contexts that the Hebbian rule neglects. By contrast, the deconvolution of the target (Equation~\ref{eq:deconv_target}) had a negligible importance ($0.001$): with the pseudo-inverse, the median loss was $0.22$ with and without it, suggesting that training compensates for the residual charge that the biphasic correction cancels at initialisation.


\paragraph{Membrane decay $\beta$.}
The membrane decay, sampled uniformly in $[0, 1]$ (as an initial value when $\beta$ was also learnt), had a modest importance ($0.02$) but a consistent effect on the robustness of training. Among the Adadelta trials, the median loss increased from $\mathcal{L} \approx 0.21$ for $\beta \le 0.2$ to $\approx 0.56$ for $0.6 < \beta \le 0.8$ and $\approx 0.73$ for $\beta > 0.8$, while the best loss reached in each of these ranges stayed between $0.09$ and $0.11$ up to $\beta = 0.8$ (best trial: $\beta \approx 0.51$, $\mathcal{L} \approx 0.087$) and degraded to $0.24$ beyond. Accordingly, the best $10\%$ of trials all had $\beta \le 0.63$ (median $0.31$). A short membrane time constant thus makes learning more tolerant to the other hyperparameters, plausibly because the residual charge left by a partial context decays before it can trigger a spurious spike, whereas a well-tuned network still performs well up to $\beta \approx 0.8$. The default value $\beta = 0.7$ ($\tau \approx 2.8$ steps) lies within this range and keeps the biphasic correction of Equation~\ref{eq:deconv_target} significant.

\paragraph{Surrogate sharpness $\alpha$.}
Within the explored range ($\alpha$ from $1.4$ to $8.1$ in the sampled trials), the sharpness of the fast-sigmoid surrogate had a negligible importance ($0.002$). Among the Adadelta trials, the median loss stayed between $0.25$ and $0.33$ and the best loss between $0.09$ and $0.11$ for $\alpha$ from $1.4$ to $5$, and the best $10\%$ of trials spread over $\alpha \in [1.5, 3.1]$ (median $1.9$). Learning is thus robust to the choice of $\alpha$ in this range, which includes the default value $\alpha = 5$; larger values were sampled too rarely ($5$ trials) to be conclusive.

\subsection{Working memory and robustness to perturbations}
\label{sec:robustness}
\begin{figure}[!t]
\centering
\includegraphics[width=0.333\linewidth]{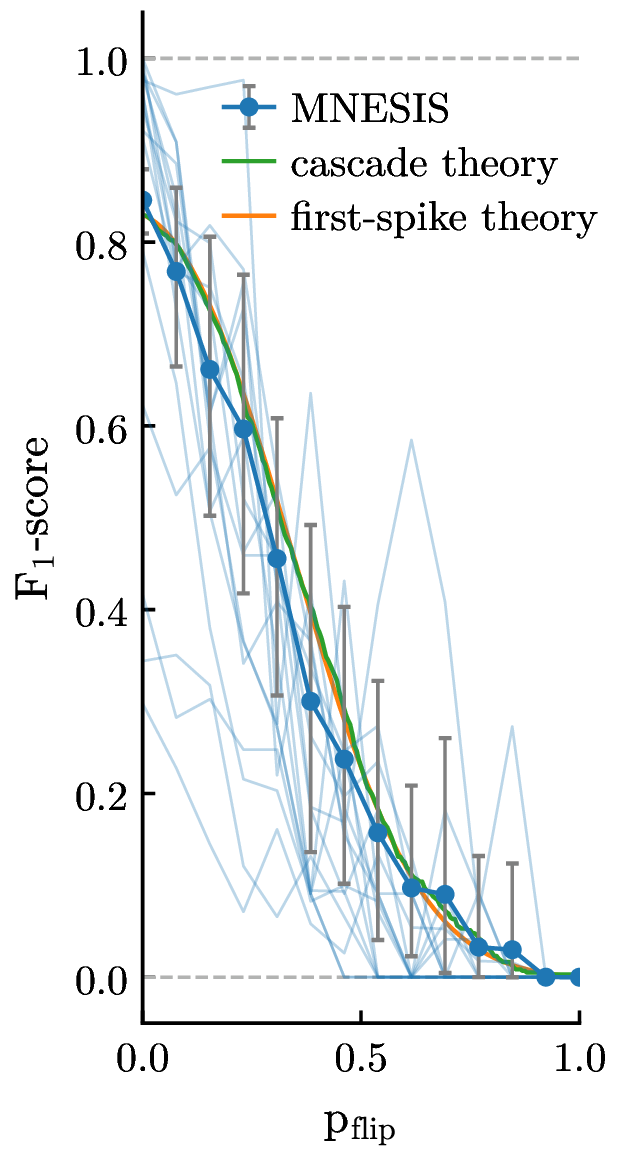}\hfill
\includegraphics[width=0.333\linewidth]{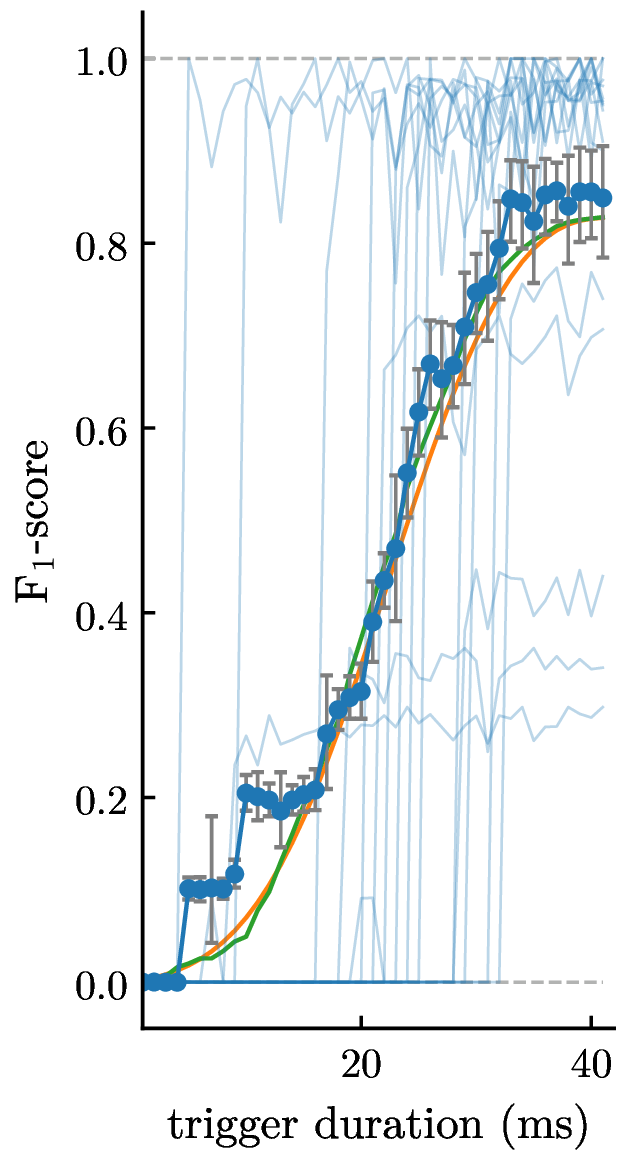}\hfill
\includegraphics[width=0.333\linewidth]{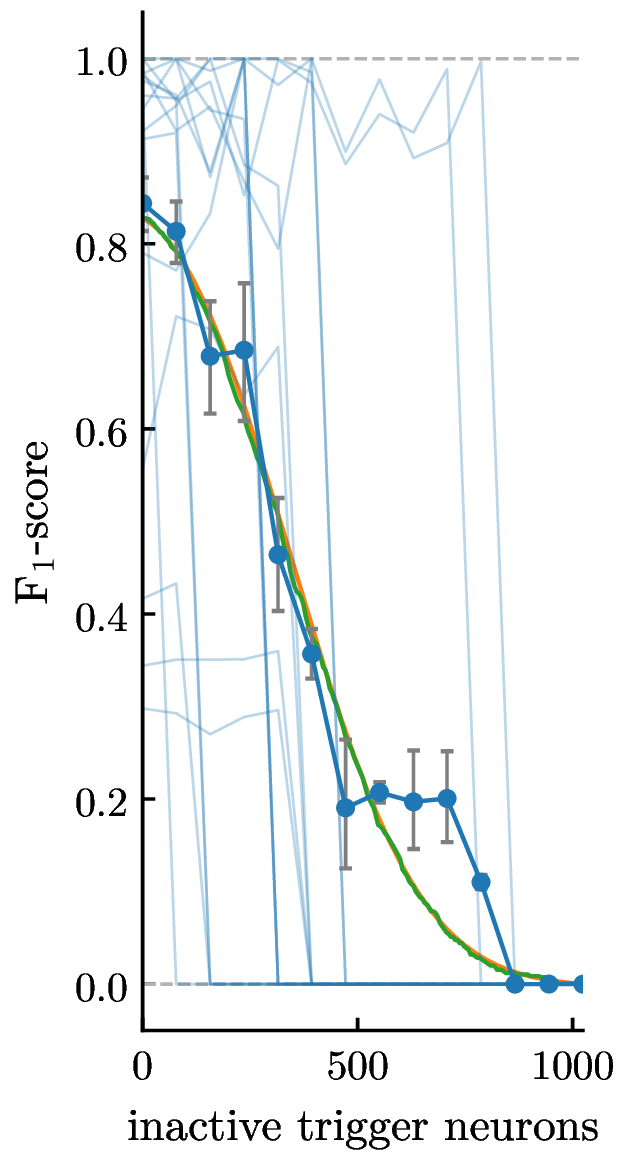}
\caption{Robustness of working-memory retrieval to three types of trigger-window perturbation. The trained MNESIS network is triggered by $11$ independent realisations of the perturbed trigger, each drawing the trigger noise and the spontaneous activity from its own random seed. The $F_1$ score is computed on the spikes of each of the $M=16$ motifs. In the plot, each point is the mean over these realisations, and error bars span the $5\%$ to $95\%$ quantiles of a beta distribution fitted to them; and thin blue lines show the $F_1$ score of each motif. \emph{Left:} Bit-flip noise applied to all spikes in the trigger window ($p_\mathrm{flip} \in [0,1]$); the network maintains high $F_1$ up to $p_\mathrm{flip} \approx 0.2$. \emph{Middle:} Truncated trigger window, reduced to its last steps (trigger duration from $1$ to $D = 41$ steps); retrieval remains reliable for durations above roughly $29$ steps. \emph{Right:} Partial neuron coverage (number of silenced neurons from $0$ to $N = 1024$); the $F_1$ score degrades smoothly as more neurons are silenced and drops sharply once the majority of the trigger population is inactive. In each panel, the orange line is the first-spike theory, the $F_1$ score expected when the replay of a motif, a cascade decided by its first evoked spike, starts only if at least a fraction $\vartheta$ of the $\approx p_A N D \approx 6.7$ spikes supporting that spike are present in the perturbed trigger window (Equation~\ref{eq:ignition}, a mean-field model of the pseudo-inverse readout, with the precision and recall of the intact trigger as its only measured inputs), and the green line is the cascade theory, the same model simulated step by step over the whole replay of $1000$ random motifs.}\label{fig:noise}
\end{figure}

%
A hallmark of biological working memory is graceful degradation: memory recall should remain possible under incomplete or corrupted retrieval triggers, mirroring the tolerance of hippocampal pattern completion to partial input~\citep{Villette2015}. Three distinct perturbations of the trigger window were therefore evaluated, each probing a different dimension of robustness (Figure~\ref{fig:noise}). The measured $F_1$ scores are first reported, then all three are accounted for with a single model of the ignition of the replay, which has no fitted parameter.

\paragraph{Robustness to spike noise (Figure~\ref{fig:noise}, left).}
The trigger window was corrupted with the balanced bit flips of Equation~\ref{eq:pattern_generation}, each bin being redrawn with probability $p_\mathrm{flip} \in [0, 1]$, and the replay was scored against the frozen motif. At $p_\mathrm{flip} = 0.25$, the network still achieved an $F_1$ score of $\approx 0.531$ (precision $\approx 0.731$, recall $\approx 0.419$), demonstrating a reasonable tolerance to noise in the trigger window. The $F_1$ score decreases monotonically as $p_\mathrm{flip}$ increases and falls below $0.1$ beyond $p_\mathrm{flip} \approx 0.6$, although the trigger window is entirely randomised only at $p_\mathrm{flip} = 1$.

\paragraph{Robustness to trigger-window duration (Figure~\ref{fig:noise}, middle).}
The trigger window was reduced to its last $L$ steps, the first $D - L$ steps carrying spontaneous activity, with a duration $L$ from $1$ to $D = 41$ steps, scanning from a single step of trigger to the full trigger window. At $75\%$ of the full trigger duration ($31$ steps out of $D = 41$), the network achieved an $F_1$ score of $\approx 0.788$ (precision $\approx 0.891$, recall $\approx 0.708$). The $F_1$ score increases monotonically with trigger duration, rising steeply once the trigger exceeds roughly $D/2$, and reaches $80\%$ of its value for the full trigger window from $L = 29$ steps.

\paragraph{Robustness to partial neuron coverage (Figure~\ref{fig:noise}, right).}
The number of neurons whose activity was silenced during the trigger window was varied, from $0$ (full trigger) to $N = 1024$ (no trigger). With $N/8 = 128$ neurons silenced --- meaning $87.5\%$ of the population remained active --- the network still achieved an $F_1$ score of $\approx 0.761$ (precision $\approx 0.849$, recall $\approx 0.690$). The $F_1$ score degrades smoothly as more neurons are silenced, drops sharply only when the majority of the trigger population is inactive, and levels off at $F_1 \approx 0.2$ beyond $n \approx N/2$ silenced neurons.

\paragraph{Replay as a cascade.}
These three behaviours follow from the way the replay unfolds: it is a cascade, not a one-shot completion of the trigger window. During the trigger window, the context of each neuron is imposed from outside. After it, the context is progressively made of the network's own output, and entirely so after $D$ steps. Each emitted spike thus joins the context of the next ones: a correct spike makes the next spikes of its motif more likely, and a missed one makes them less likely. This feedback loop has two stable states. Either the evoked spikes complete the contexts of the next ones, and the replay ignites and sustains itself over the whole motif, the noise-free motif being a stable fixed point of the recurrent dynamics; or they fall short, the contexts keep shrinking, and the replay dies out. A perturbation of the trigger window therefore does not degrade the replay gradually: it decides, motif by motif, which state is reached, so that the $F_1$ score measures the fraction of motifs that ignite. In this sense, the network operates as an attractor: small perturbations are corrected by the recurrent dynamics, which converge back to the stored motif, as long as a sufficient part of the motif remains to ignite it. Two observations support this picture. First, for the deterministic perturbations, the truncation of the trigger window and the lesion of a fixed set of neurons, the $F_1$ score of each motif (thin lines in Figure~\ref{fig:noise}) switches abruptly between zero and its intact value, each motif at its own level of perturbation; under random flips, where it is averaged over the realisations of the noise, it decays smoothly, as the probability that the motif ignites. Second, in an open-loop control in which the output spikes are not fed back to the network, the trigger window alone drives the first $D$ steps after it, after which the activity vanishes: the replay is sustained by the recurrent feedback, not by the trigger.

The cascade is decided at its first step: the first evoked spike of a motif either sets off the positive feedback loop or does not. The first-spike model computes this decision in closed form, starting from the analytical initialisation (Equation~\ref{eq:init}) in a mean-field approximation. The minimum-norm solution gives equal weights to the $K$ spikes of a motif that lie in the $D$ steps before one of its spikes, so that the membrane potential reaches $1$ when all of them are present and $k/K$ when only $k$ of them are, while a spurious spike of the trigger window falls on a near-zero weight, the motifs being sparse. The spike is thus emitted if $k \geq \lceil \vartheta K \rceil$. The probability $R$ that a motif ignites combines three named factors:
\begin{enumerate}
  \item \emph{Timing (Poisson approximation).} A motif emits $\nu = p_A N \approx 0.16$ spikes per step. Approximating these spikes as a Poisson process of rate $\nu$, the first evoked spike occurs $\Delta$ steps after the trigger window with probability $P(\Delta) = e^{-\nu \Delta}(1 - e^{-\nu})$.
  \item \emph{Context size (Poisson count).} The context of this spike is made of the last $D - \Delta$ steps of the trigger window. It holds a Poisson number $K$ of spikes, of mean $\lambda_\Delta = \nu (D - \Delta)$, that is $p_A N D \approx 6.7$ for $\Delta = 0$.
  \item \emph{Ignition (threshold and binomial thinning).} Each of these $K$ spikes is still present in the perturbed trigger window with probability $q_\Delta$, the presence probability. The spike is emitted if at least $\lceil \vartheta K \rceil$ of them are present.
\end{enumerate}
Combining the three factors gives
\begin{equation}
  R = \frac{1}{R_0} \sum_{\Delta=0}^{D-1} \underbrace{P(\Delta)}_{\text{timing}} \sum_{K \geq 1} \underbrace{\mathrm{Poi}(K;\lambda_\Delta)}_{\text{context size}} \underbrace{\sum_{k = \lceil \vartheta K \rceil}^{K} \mathrm{Bin}(k;K,q_\Delta)}_{\text{ignition}},
  \label{eq:ignition}
\end{equation}
where $\mathrm{Poi}(K;\lambda) = e^{-\lambda}\lambda^K/K!$ is the Poisson probability and $\mathrm{Bin}(k;K,q) = \binom{K}{k} q^k (1 - q)^{K-k}$ the binomial one. The normalisation $R_0 \approx 0.991$ is the value of the double sum for the intact trigger window ($q_\Delta = 1$), so that $R = 1$ for an intact trigger; the remaining failures are those of the motifs whose first evoked spike follows a gap (Section~\ref{sec:task}), already counted in the recall of the intact trigger. An ignited motif is replayed with the precision $\pi_0$ and recall $r_0$ measured with the intact trigger window, and a silent one only adds false negatives, so that $F_1 = 2 R\, r_0 / (1 + R\, r_0 / \pi_0)$. The three perturbations differ only by the presence probability
\begin{equation*}
  q_\Delta = \big(1 - p_\mathrm{flip}(1 - p_A)\big) \times
  \begin{cases}
    1 & \text{spike noise ($p_\mathrm{flip}$ scanned)},\\[2pt]
    \dfrac{\min(L, D - \Delta)}{D - \Delta} & \text{trigger duration $L$},\\[8pt]
    1 - n/N & \text{$n$ silenced neurons},
  \end{cases}
\end{equation*}
with $p_\mathrm{flip} = 0.01$ for the last two. The first factor is the probability that a spike of the trigger window survives the flips. Silencing $n$ neurons removes the same fraction of every part of the window, so that, as for the flips, $q_\Delta$ does not depend on $\Delta$. The truncated trigger, in contrast, keeps the most recent part of the window, where the context of the first evoked spike lies, so that $q_\Delta$ is larger than the share $L / D$ of the window.

This theory, the ignition of the replay by its first evoked spike, is compared at two levels of approximation with the measured $F_1$ scores (Figure~\ref{fig:noise}). The first-spike theory (orange lines) is Equation~\ref{eq:ignition}, which computes in closed form only the first step of the cascade. The cascade theory (green lines) simulates the whole replay instead, step by step, on $1000$ random motifs of the same statistics and duration $T$, the emitted spikes joining the context of the next ones and the neurons keeping their identity in the trigger window; its recall, relative to that of the intact trigger window, gives $R$. The two theories differ by at most $0.04$ on the three perturbations, which confirms that the first evoked spike decides the replay, and both follow the measured $F_1$ scores with no fitted parameter: the RMS error is $0.042$ for both theories under spike noise, $0.055$ (first-spike) and $0.052$ (cascade) for the truncated trigger, and $0.061$ for both under the lesion. The variability, from one motif to the next, of the number of spikes that support the first evoked spike also accounts for each motif having its own sensitivity to the perturbation (thin lines). Two deviations remain. For the truncated trigger, unlike the two other perturbations, later evoked spikes, whose context is even more recent, give a motif a few more chances to ignite when the first one fails, so that the cascade theory lies slightly above the first-spike theory. For the lesion, the plateau at $F_1 \approx 0.2$ that lies above both theories beyond $n \approx N/2$ is a finite-size effect, as the lesion always silences the same addresses and the two of the $M = 16$ motifs whose trigger spikes fall mostly on the remaining neurons keep igniting.

\subsection{Retrieval over long horizons: chains of periodic spike motifs}
\label{sec:periodic}
%
\begin{figure}[!t]
\centering
\includegraphics[width=\linewidth]{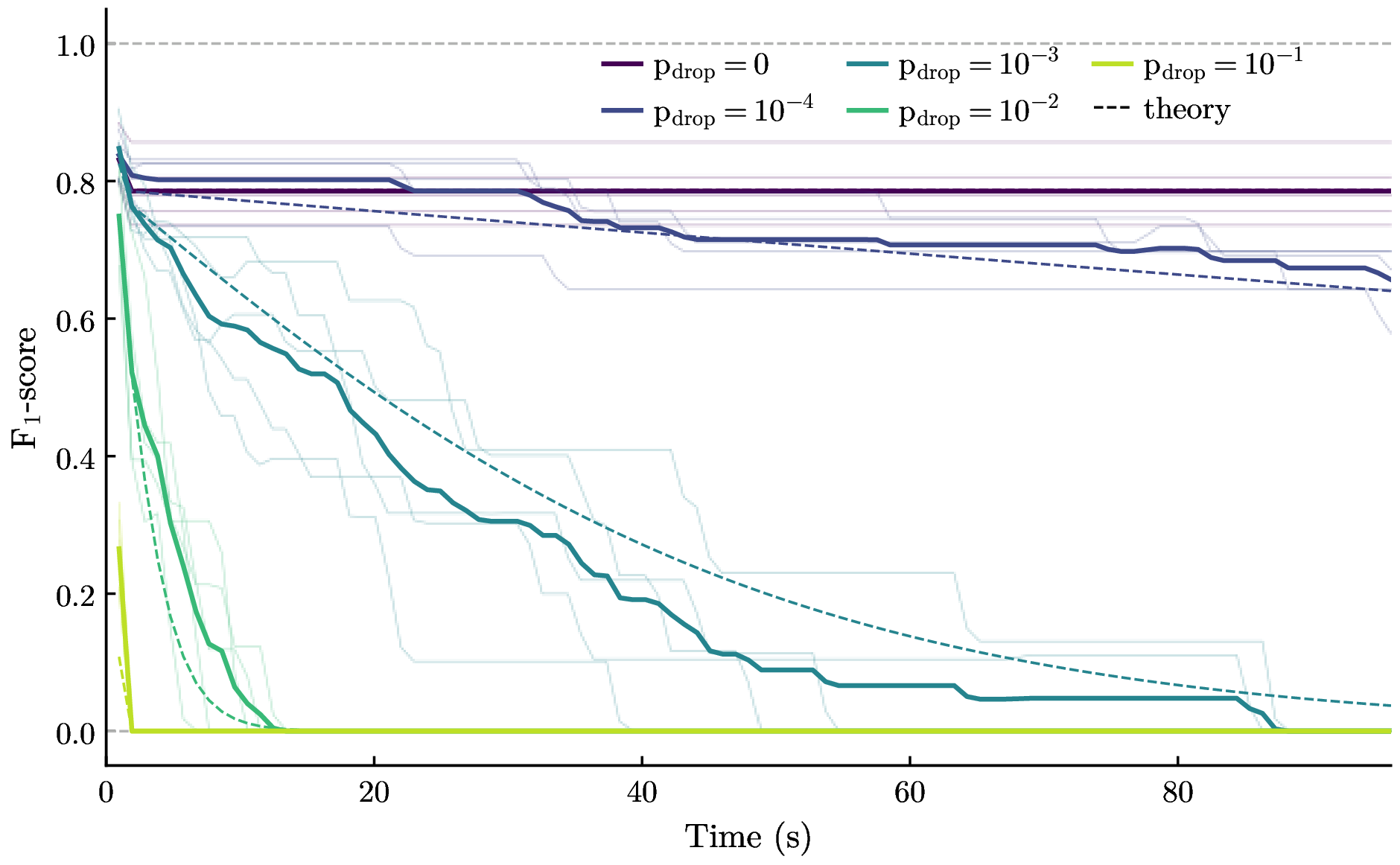}
\caption{Free replay of periodic motifs under transmission failures. The motifs are made periodic by copying their trigger window (the first $D=41$ steps) onto their last $D$ steps, so that the chain of Spiking Motifs closes on itself with a period of $T-D=959$ steps of $1\,\mathrm{ms}$, that is $0.959\,\mathrm{s}$, about one second. After a single noisy trigger window ($p_\mathrm{flip} = 0.01$), each spike emitted by the network is removed from the recurrent input with probability $p_\mathrm{drop}$, and the network runs freely for $100$ periods ($95.9\,\mathrm{s}$). Each curve is one level of $p_\mathrm{drop}$, as given by the legend: $p_\mathrm{drop} = 0$ and four log-spaced levels from $10^{-4}$ to $10^{-2}$. Thin lines show the $5$ seeded realisations of each level, thick lines their mean. Each period is scored ($F_1$ score computed on the spikes of the $M=16$ motifs pooled together) against the frozen motif, and plotted at the time at which the period ends, counted from the end of the trigger window. Dashed lines: prediction of the cascade model (Section~\ref{sec:robustness}), in which each spike of a sustained replay is a new ignition test, so that the replay survives $t$ steps with probability $\exp(-ht)$. Without failures, the score settles on the limit cycle after the first period; with failures, the replay decays with a mean lifetime $1/(h(T-D))$ of about $280$, $60$, $12$ and $2$ periods at the four non-zero levels.}\label{fig:periodic}
\end{figure}

%
The working memory task above has a natural prolongation to long horizons. As it stands, it has an end: once the last context window of a motif has been replayed, nothing in the stored chain predicts further activity, and the network falls silent. A simple modification of the motifs turns this finite chain into an endless one. For each frozen motif, its trigger window was copied onto its last $D$ time steps, that is, $P^{(\mu)}[\cdot,\, T-D+t] = P^{(\mu)}[\cdot,\, t]$ for $0 \le t < D$, the same wrap being applied to each noisy realisation after the bit flips of Equation~\ref{eq:pattern_generation}. The last context window of a motif is then identical to its trigger window, so that a faithful replay of the end of the motif re-triggers its beginning: the chain of overlapping Spiking Motifs closes on itself, with a period of $T - D = 959$ steps of $1\,\mathrm{ms}$, that is $0.959\,\mathrm{s}$, about one second. The network was trained exactly as in the main experiment, with the same parameters and on a single period of $T = 1000$ steps, so that it never experienced the transition from one period to the next during learning. Triggered by a noisy realisation of the trigger window ($p_\mathrm{flip} = 0.01$), the replay of the first period reached an $F_1$ score of $\approx 0.827$ (precision $\approx 0.828$, recall $\approx 0.826$).

The network was then triggered with the noise-free trigger window and run freely for $30$ periods ($28\,770$ steps, about $29\,\mathrm{s}$), each period being scored separately against the frozen motif. The first period reached an $F_1$ score of $\approx 0.880$ (precision $\approx 0.940$, recall $\approx 0.827$), and from the second period onwards the score stayed at $\approx 0.829$ (precision $\approx 0.934$, recall $\approx 0.746$) without any further decline. The evoked rasters were strictly identical from the second period onwards: the network settled into an exact limit cycle, whereas the first period differs from the second. The drop in recall between the first two periods reflects an all-or-none selection at the first closure of the cycle. For $12$ of the $M = 16$ motifs, the replay reached the end of the motif and re-triggered itself, after which it was sustained indefinitely; for the $4$ remaining motifs, the replay had already faded within the first period (from $28$ to $92$ evoked spikes) and these motifs stayed silent afterwards. These $4$ motifs are precisely those that contain a gap of $D$ silent steps (Section~\ref{sec:task}), which the wrap leaves untouched: their replay stops at the onset of the gap, whatever the trigger, and cannot reach the end of the motif to re-trigger its beginning. The stationarity of the score follows from deterministic recurrent dynamics: once the activity over the last $D$ steps of a period matches that of the previous period, the next period is replayed identically. As a consequence, errors do not accumulate from one period to the next; they are settled once and for all during the first period.

To test how long a periodic memory lasts when its replay is imperfect, this long-horizon replay was combined with transmission failures. After the noisy trigger window of Section~\ref{sec:robustness} ($p_\mathrm{flip} = 0.01$), each spike emitted by the network fails to reach the recurrent input with probability $p_\mathrm{drop}$, independently at each step, and the network runs freely for $100$ periods ($95.9\,\mathrm{s}$, that is about one and a half minutes of biological time), each period being scored against the frozen motif (Figure~\ref{fig:periodic}). Five levels were used, $p_\mathrm{drop} = 0$ and four log-spaced values from $10^{-4}$ to $10^{-2}$, each on $5$ seeded realisations. Each spike of a sustained replay is a new ignition test, so that the replay survives $t$ steps with probability $e^{-ht}$, where the failure rate $h$ grows with $p_\mathrm{drop}$; the predicted $F_1$ score uses the ignition theory of Section~\ref{sec:robustness}, with the precision and recall of the replay without failures as its only measured inputs. Without failures, the replay settles on its limit cycle: the $F_1$ score is $0.831 \pm 0.042$ in the first period and $0.785 \pm 0.058$ in the tenth (mean $\pm$ 95\,\% confidence interval over the $5$ realisations, Student $t$ with $4$ degrees of freedom; the theory gives $0.833$ and $0.787$). The first period is almost unaffected up to $p_\mathrm{drop} = 2.2\times10^{-3}$ (between $0.84$ and $0.85$), and the later decay follows the survival law at every level, within the confidence intervals. The mean lifetime $1/(h(T-D))$ of a replay falls from $\approx 280$ periods at $p_\mathrm{drop} = 10^{-4}$ to $\approx 60$ at $4.6\times10^{-4}$, $\approx 12$ at $2.2\times10^{-3}$ and $\approx 2$ at $10^{-2}$. At $10^{-4}$ the score is $0.657 \pm 0.062$ at period $100$ (theory $0.640$); at $4.6\times10^{-4}$ it is $0.413 \pm 0.064$ at period $50$ (theory $0.465$) and $0.274 \pm 0.072$ at period $100$ (theory $0.238$); at $2.2\times10^{-3}$ it is $0.024 \pm 0.066$ at period $50$ (theory $0.025$); at $10^{-2}$ the replay vanishes within about ten periods ($0.042 \pm 0.073$ at period $10$, theory $0.018$). The no-failure level is lower than the noise-free run above ($0.829$), because its trigger is noisy; with failures, the decay rate is set by the failure rate alone.

%
\subsection{Effect of delay depth, network size, motif duration and number of motifs}
\begin{figure}[!t] 
  \centering
  \includegraphics[width=0.5\linewidth]{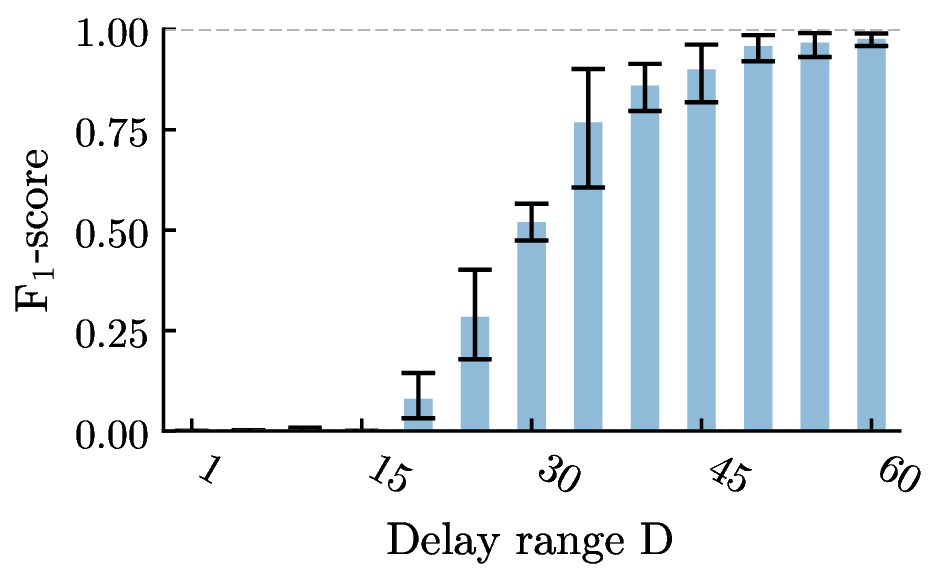}%
  \includegraphics[width=0.5\linewidth]{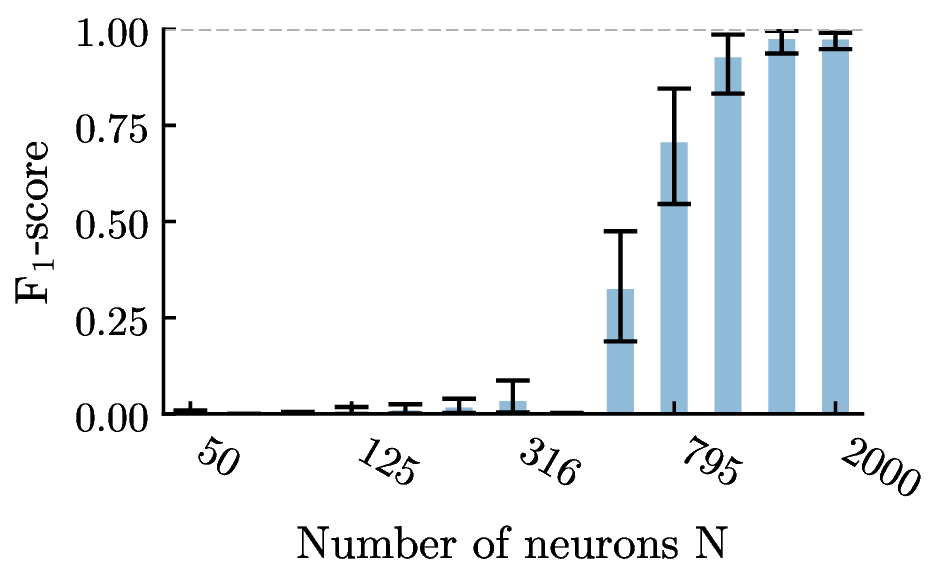}\\
  \includegraphics[width=0.5\linewidth]{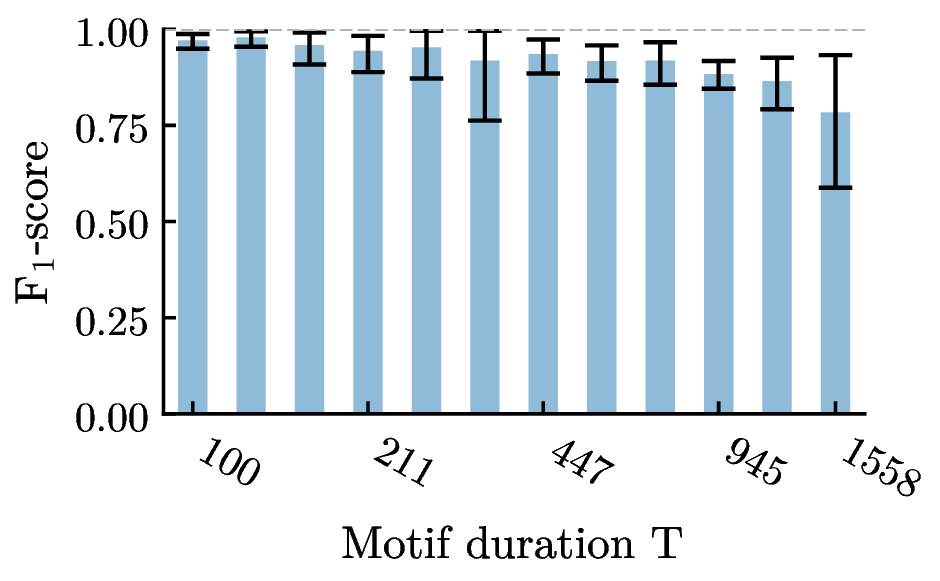}%
  \includegraphics[width=0.5\linewidth]{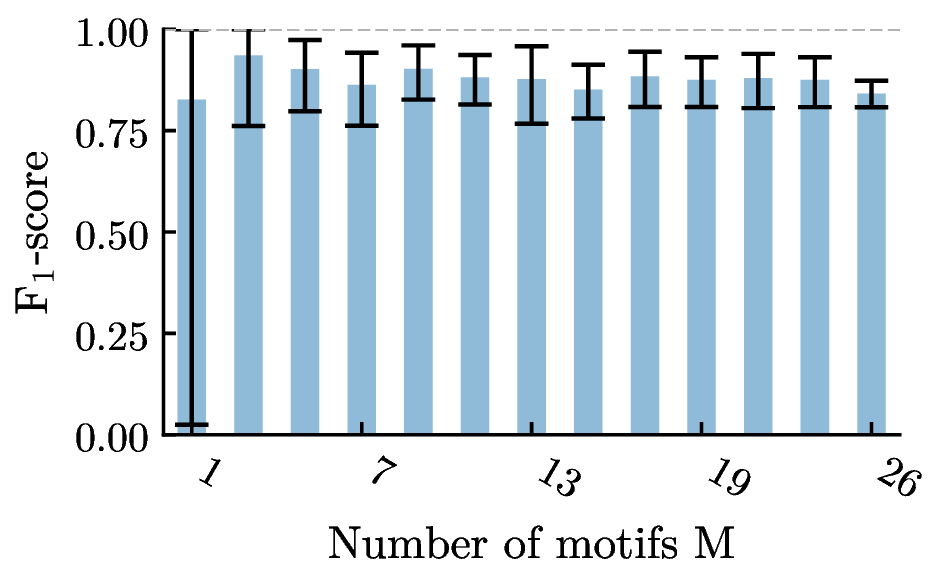}
  \caption{%
    Effect of some key parameters on working-memory performance ($F_1$ score over $10$ independent runs, each training a new network on new motifs drawn from its own random seed, a given seed sharing its motifs across all values of a parameter; bars show the mean over these runs, and whiskers span the $5\%$ to $95\%$ quantiles of a beta distribution fitted to them). %
    \emph{Top:}
    The $F_1$ score rises steeply with the maximum delay $D$ (left) and with the number of neurons $N$ (right): both increase the mean number $p_A N D$ of spikes in a context window, and thus decrease the probability that a motif contains a gap of $D$ silent steps, at which its replay stops. %
    \emph{Bottom:}
    It decreases slowly with the motif duration $T$ (left), as a longer motif is more likely to contain such a gap, and does not depend on the number of stored motifs $M$ over the scanned range, up to $M = 26$ (right).
  }\label{fig:scan}
\end{figure}

To characterise how the architectural parameters govern working-memory performance, a systematic one-at-a-time scan was conducted over the maximum delay $D$, the number of neurons $N$, the motif duration $T$ and the number of stored motifs $M$, holding all other hyperparameters at their default values (Section~\ref{sec:methods}). Each condition was evaluated with $10$ independent runs (random seeds) and results are reported as the mean $F_1$ score (Section~\ref{sec:task}) of the evoked window (Figure~\ref{fig:scan}). These scores are compared with the recall that the cascade model of Section~\ref{sec:robustness} predicts for an intact trigger window, simulated on random motifs of the same statistics: with no fitted parameter, it accounts for the failures at a gap of $D$ silent steps (Section~\ref{sec:task}), and leaves out the crosstalk between motifs, which lowers the precision.

\paragraph{Delay depth $D$ (Figure~\ref{fig:scan}, top left).}
The $F_1$ score is close to zero for $D \leq 15$, rises steeply from $\approx 0.08$ at $D = 20$ to $\approx 0.86$ at $D = 40$, and exceeds $0.95$ from $D = 50$ onwards ($\approx 0.98$ at $D = 60$). The cascade model predicts this threshold: a motif holds on average $(T-D)(1 - e^{-p_A N})\, e^{-p_A N D}$ gaps (Section~\ref{sec:task}), a number that falls exponentially with the mean number $\lambda_0 = p_A N D$ of spikes in a context window, so that the predicted recall rises from $0.15$ at $D = 20$ to $0.89$ at $D = 40$ and $1.00$ at $D = 60$, within $0.1$ of the measured $F_1$ score over the whole range. A longer delay range thus does not act through the number of weights, but by making each context window rich enough to support the next spike of the motif.

\paragraph{Network size $N$ (Figure~\ref{fig:scan}, top right).}
The $F_1$ score remains close to zero up to $N \approx 430$ neurons, then rises steeply to $\approx 0.33$ at $N = 584$, $\approx 0.7$ at $N = 795$ and $\approx 0.93$ at $N = 1081$, and stays high ($\approx 0.97$) from $N \approx 1500$ onwards. The number of neurons enters the gap probability through the same quantity $\lambda_0 = p_A N D$ as the delay range, and the two scans agree once expressed in $\lambda_0$: $N = 584$ ($\lambda_0 \approx 3.8$, $F_1 \approx 0.33$) behaves as $D = 25$ ($\lambda_0 \approx 4.1$, $F_1 \approx 0.28$). The cascade model again follows the measured scores within $0.1$, except at $N = 430$, where it predicts a recall of $0.19$ while the network stays silent.

\paragraph{Motif duration $T$ (Figure~\ref{fig:scan}, bottom left).}
The $F_1$ score decreases slowly with $T$, from $\approx 0.97$ at $T = 100$ to $\approx 0.88$ at $T = 945$ and $\approx 0.78$ at $T = 1558$. A longer motif is more likely to contain a gap, as their number grows linearly with $T - D$, and the cascade model predicts a recall decreasing from $0.98$ to $0.90$ and $0.87$ over the same range. The faster decrease of the measured score at the longest durations points to the crosstalk between motifs, which the model leaves out.

\paragraph{Number of motifs $M$ (Figure~\ref{fig:scan}, bottom right).}
The $F_1$ score does not depend on the number of stored motifs: it lies between $0.84$ and $0.94$, with no trend, from $M = 3$ to $M = 26$. A gap is a property of each motif, independent of the others, so that the fraction of motifs replayed to their end does not change with $M$, and neither does the score. With a single motif ($M = 1$), the mean score ($\approx 0.83$) is the most variable, as a run whose motif holds an early gap scores close to zero ($0.06$ for one of the $10$ runs). At $M = 26$, each neuron must map $M (T - D) \approx 25{,}000$ context windows to its next state with its $N D \approx 42{,}000$ incoming weights, a load of $\alpha = M(T-D)/(ND) \approx 0.59$, and the replay still shows no sign of interference.


Taken together, the scans show that the replay is limited by the statistics of each motif rather than by the number of stored motifs: the delay range $D$ and the population size $N$ act through the mean number $\lambda_0 = p_A N D$ of spikes in a context window, which sets the probability of a gap, the duration $T$ through the number of context windows in which a gap may occur, and the number of motifs $M$ not at all over the scanned range. Increasing $D$ is the most economical way to raise $\lambda_0$: doubling $D$ doubles the number $N^2 D$ of weights, whereas doubling $N$ quadruples it and adds as many neurons. Increasing $D$ is, however, ultimately limited by biological constraints on axonal conduction delays~\citep{Grimaldi2023a}.

\section{Discussion}

\subsection{\ShortTitle}

The results show that the MNESIS recurrent spiking neural network with heterogeneous synaptic delays (HD-SNN) can store arbitrary spike motifs and replay them from a brief and noisy trigger window on a benchmark of $M=16$ motifs of $T=1000$ steps using chains of polychronous groups. What is remarkable is not the number of weights, $N^2 D \approx 43 \times 10^6$, which far exceeds the few thousand spikes of the stored motifs, but the number of context windows that these weights keep apart. Each neuron must map each of the $M(T-D) \approx 15{,}000$ context windows of the motifs to its next state with its $N D \approx 42{,}000$ incoming weights. These windows are moreover almost empty, with $\lambda_0 = p_A N D \approx 6.7$ spikes on average among their $N D$ entries, and two successive windows share $D - 1$ of their $D$ steps. They are nevertheless distinct enough for the network to chain them over the whole motif, as a sequence of overlapping Spiking Motifs~\citep{Perrinet2023,Kronland-Martinet2026}, and the scan of the number of motifs shows no interference up to $M = 26$, that is, $\approx 25{,}000$ context windows per neuron.

The parameter scans identify the delays as the critical parameter that makes this storage possible. Indeed, the number of neurons and the average firing rate influence the metabolic cost and should be strictly bounded~\citep{Lennie2003}. Numerically, the replay fails for $D \leq 15$ and becomes reliable from $D \approx 50$, and at a comparable load $\alpha = M(T-D)/(ND) \approx 0.6$, the score depends on the context window rather than on the number of motifs: $F_1 \approx 0.84$ with $M = 26$ motifs, but $0.28$ with a delay range $D = 25$ and $0.33$ with $N = 584$ neurons. What limits the replay is the number $\lambda_0$ of spikes that support each evoked spike, which the cascade model turns, with no fitted parameter, into the probability of reaching the end of the motif. A longer delay range raises $\lambda_0$ linearly without adding any neuron, and at a lower cost in weights than a larger population. This provides concrete evidence that heterogeneous delays are a computational asset rather than a biological nuisance, a point argued theoretically in~\citet{Izhikevich2006} and demonstrated empirically in the context of learnable delays in feedforward~\citep{Grimaldi2023a,Hammouamri2023} and recurrent architectures~\citep{Queant2025,Meszaros2025}, and and is extended here to a full working-memory setting.

The robustness experiments further strengthen this interpretation. Retrieval remains well above chance with $25\%$ bit-flip noise in the trigger window ($F_1 \approx 0.53$), becomes reliable once the trigger exceeds roughly $29$ of its $D = 41$ steps, and still reaches $F_1 \approx 0.76$ when $128$ of the $N = 1024$ trigger neurons ($12.5\%$) are silenced. This progressive degradation profile is consistent with the network operating as an attractor: the stored motifs correspond to stable fixed points of the recurrent dynamics, and the trigger window provides a partial trigger that the network completes autonomously; the ignition model shows that this completion is a cascade, a positive feedback loop decided motif by motif by the few trigger spikes that support the first evoked spike, and the open-loop control shows that the replay is then sustained by the recurrent feedback rather than by the trigger --- the same mechanism proposed for hippocampal pattern completion~\citep{Villette2015}.

\subsection{Limitations and directions for improvement}

The present model is deliberately minimal, and several design choices impose limits on capacity and robustness that future work should address.

\paragraph{Capacity and scaling.}
A systematic evaluation of memory capacity --- how many motifs of what duration can be reliably stored and retrieved --- remains to be carried out beyond the present scan, which showed no interference up to $M = 26$ motifs (a load $\alpha \approx 0.59$). The pseudo-inverse of the analytical initialisation fits the motifs exactly as long as the linear system of each neuron stays underdetermined, that is, up to $\alpha = 1$, or $M \approx N D / (T - D) \approx 44$ motifs of duration $T = 1000$. The threshold of the neurons and the sparseness of the motifs should allow more, as for a perceptron with sparse outputs, but whether the network reaches such a bound at scale is an open empirical question, as the implementation does not scale well with the memory needed to store the weight tensor. Interference between stored motifs at high memory load may degrade precision in ways that the F1 training objective does not fully penalise, and the sensitivity of recall to perturbations of the trigger window is expected to decrease as $M$ grows. A further limitation is the generality of the benchmark. The motifs are drawn independently, as Bernoulli rasters with no shared structure: each is a random point in a space of dimension $NT$, which makes them the hardest to compress and leaves nothing to generalise from one motif to the next. Spiking activity in the brain is, by contrast, highly structured, population activity being confined to low-dimensional manifolds~\citep{Gallego2017}. Measuring capacity and scaling in a way that transfers to neural data will require a more general generative model, for instance motifs generated by low-dimensional latent dynamics or composed of shared sub-motifs, against which one could test not only the storage of the trained motifs but also the replay of new motifs of the same family.

\paragraph{Sparse connectivity.}
In all generality, the weight tensor $\mathbf{W}$ is dense, but the minimum-norm solution of Equation~\ref{eq:init} concentrates its largest weights on the delayed coincidences between a spike of a motif and the spikes of its context window, so that each stored spike is supported by a handful of synapses with matching delays, as in a polychronous group~\citep{Izhikevich2006}. Pruning the weakest synapses at inference, as suggested in Section~\ref{sec:methods}, could thus substantially reduce the memory and energy footprint of the network, and the number of synapses kept per stored spike would measure the cost of a memory in terms of connectivity. This could be further enhanced by introducing a regularisation in line with sparse coding~\citep{Perrinet2002image}. Whether the replay, and its robustness to a noisy trigger window, survive such pruning remains to be tested.

\paragraph{Online and unsupervised learning.}
The present model requires a fixed set of target motifs known in advance. A natural extension is an online learning rule that updates the weights continuously as new motifs are presented, without catastrophic forgetting of previously stored sequences. The closed-form Hebbian initialisation (Equation~\ref{eq:init}) suggests a natural online update: each new spike $s_j(t)$ incrementally adjusts $w_{i, j, d}$ in proportion to $s_i(t-d)$, recovering a form of local STDP-like rule~\citep{Izhikevich2006}. Whether such online updates are stable under recurrent feedback and can support continual working memory remains to be established. Beyond such local rules, the training itself could be made online: the present refinement relies on backpropagation through time with a surrogate gradient (Section~\ref{sec:methods}), whereas recent methods train spiking networks with smoothed delays online, with temporally precise spikes and no surrogate gradient~\citep{Koopman2026}. Such methods could learn the weights, and possibly the delays themselves, as the motifs are presented, a prerequisite for a working memory that is written in real time.

\paragraph{Latent reservoir neurons.}
The current architecture connects every neuron directly to every other neuron across all delays. Partitioning the population into input-output neurons and a larger pool of latent ``reservoir'' neurons, as in delay-based reservoir computing~\citep{Paugam-Moisy2008}, would enrich the internal spiking dynamics without increasing the number of trained parameters proportionally. Such a hybrid architecture could combine the representational power of random recurrent networks with the precision of gradient-trained delay weights~\citep{Neftci2019a}, potentially closing the gap between biologically inspired models and sparse selective-update architectures~\citep{Yin2026}. It is also possible to integrate the concept of chunking in a hierarchical architecture~\citep{Zhong2026}, where the recurrent HD-SNN serves as a high-level controller that orchestrates the activation of multiple parallel reservoirs, each specialised for different temporal scales or motif types.

\paragraph{Richer neuron models and dynamics.}
The simplest possible spiking neuron model is used, a LIF with a fixed membrane decay $\beta$. Introducing adaptive threshold dynamics or more complex single-neuron models, as studied in~\citet{Szatmary2010}, should increase the diversity of intrinsic timescales available to the network and improve both capacity and retrieval fidelity. In particular, the deconvolution analysis (Section~\ref{sec:methods}) shows that the optimal weight profile depends critically on $\beta$: larger values of $\beta$ require exponentially amplified weights at long delays, suggesting that adaptive $\beta$ per neuron could allow the network to exploit the full delay range more efficiently. Likewise, the synapses are current-based, so that the effect of a presynaptic spike does not depend on the state of the postsynaptic neuron. Conductance-based synapses could make this effect multiplicative: a modulatory input, by changing the conductance of a neuron, could gate or rescale the replay of a motif without changing the stored weights~\citep{Boutin2020}. Such multiplicative interactions are a natural substrate for hierarchies of motifs, in which a higher level selects which lower-level motifs are replayed, as in the chunking architecture discussed above.

\subsection{Perspectives: unsupervised discovery of spiking motifs in neural data}
It is hypothesised that the spiking activity of neural populations is organised into Spiking Motifs, and that MNESIS is a reference model for testing this hypothesis. The most immediate application of this framework is to neurophysiological data. The supervised memorisation task studied here requires a known target motif, but in practice one would like to discover the latent Spiking Motifs present in recorded multi-unit activity --- the problem originally addressed by~\citet{Ikegaya2004} and~\citet{Villette2015}.

Applied to neural data, the model can analyse recordings and generate hypotheses on the structure of Spiking Motifs. A natural extension is to replace the supervised F1 loss with a self-supervised or contrastive objective that trains the network to predict upcoming spikes from the recent spike history, without access to ground-truth labels. This is analogous to the contrastive predictive coding paradigm applied to continuous neural time series, here instantiated in a fully spike-based setting~\citep{Grimaldi2023}. Such a model would provide a principled, energy-efficient method for identifying the temporal structure of neural population codes, with direct relevance to the debate on rate coding versus temporal coding~\citep{Softky1993,Bohte2004} and to understanding the neural mechanisms of working memory more broadly.

If the hypothesis holds, MNESIS also offers an energy-efficient way to test it on neuromorphic hardware, which matters given the current profusion of neural data. Deployed on neuromorphic hardware~\citep{Goltz2021,Eshraghian2023}, it could operate in real time on streaming electrophysiological data, opening a new avenue for closed-loop brain-machine interface applications.
%
\section*{Acknowledgements}
\noindent \Acknowledgements
\section*{Statement on the usage of Generative Artificial Intelligence}
\noindent \AIstatement
\printbibliography
\end{document}